# The Jade Gateway to Trust: Exploring How Socio-Cultural Perspectives Shape Trust Within Chinese NFT Communities

YI-FAN CAO, Hong Kong University of Science and Technology, China
REZA HADI MOGAVI, University of Waterloo, Canada
MENG XIA, Texas A&M University, USA
LEO YU-HO LO, Hong Kong University of Science and Technology, China
XIAO-QING ZHANG, University of International Business and Economics, China
MEI-JIA LOU, Hong Kong University of Science and Technology, China
LENNART E. NACKE, University of Waterloo, Canada
YANG WANG, Hong Kong University of Science and Technology, China
HUAMIN QU, Hong Kong University of Science and Technology, China

Today's world is witnessing an unparalleled rate of technological transformation. The emergence of non-fungible tokens (NFTs) has transformed how we handle digital assets and value. These tokens have captured the interest of scholars and businesspeople alike. However, NFTs have recently seen a sharp decline in popularity. While cryptocurrency volatility and monetary policies greatly influenced NFT market trends, the community aspects of NFT projects–particularly trust-based interactions–also play a crucial role in NFT adoption and sustainability. From a social computing perspective, understanding these trust dynamics offers valuable insights for the development of both the NFT ecosystem and the broader digital economy. China presents a compelling context for examining these dynamics, offering a unique intersection of technological innovation and traditional cultural values. Through an in-depth qualitative study of Chinese NFT communities, we examine how socio-cultural factors influence trust formation and development. We analyzed discussions from eight prominent WeChat groups dedicated to NFTs and conducted 21 semi-structured interviews with three types of NFT community members. We found that trust in Chinese NFT communities is significantly molded by local cultural values. To be precise, Confucian virtues, such as *benevolence*, *propriety*, and *integrity*, play a crucial role in shaping these trust relationships. Our research identifies three critical trust dimensions in China's NFT market: (1) *technological*, (2) *institutional*, and (3) *social*. We examined the challenges in cultivating each dimension. Based on these insights, we developed tailored trust-building guidelines for Chinese NFT stakeholders. These guidelines address trust issues that factor into NFT's declining popularity and could offer valuable strategies for CSCW researchers, developers, and designers aiming to enhance trust in global NFT communities. Our research urges CSCW scholars to take into account the unique socio-cultural contexts when developing trust-enhancing strategies for digital innovations and online interactions.

Authors' Contact Information: Yi-Fan Cao, Hong Kong University of Science and Technology, Hong Kong, China, ycaoaw@connect.ust.hk; Reza Hadi Mogavi, University of Waterloo, Waterloo, Ontario, Canada, rhadimog@uwaterloo.ca; Meng Xia, Texas A&M University, College Station, Texas, USA, mengxia@tamu.edu; Leo Yu-Ho Lo, Hong Kong University of Science and Technology, Hong Kong, China, yhload@cse.ust.hk; Xiao-Qing Zhang, University of International Business and Economics, Beijing, China, 202320511501@uibe.edu.cn; Mei-Jia Lou, Hong Kong University of Science and Technology, Hong Kong, China, mlouaa@connect.ust.hk; Lennart E. Nacke, University of Waterloo, Waterloo, Ontario, Canada, lennart.nacke@uwaterloo.ca; Yang Wang, Hong Kong University of Science and Technology, Hong Kong, China, yangwang@ust.hk; Huamin Qu, Hong Kong University of Science and Technology, Hong Kong, China, huamin@ust.hk.

**ACM Reference Format:**
Yi-Fan Cao, Reza Hadi Mogavi, Meng Xia, Leo Yu-Ho Lo, Xiao-Qing Zhang, Mei-Jia Lou, Lennart E. Nacke, Yang Wang, and Huamin Qu. 2025. The Jade Gateway to Trust: Exploring How Socio-Cultural Perspectives Shape Trust Within Chinese NFT Communities. *Proc. ACM Hum.-Comput. Interact.* 9, 2, Article CSCW189 (April 2025), 39 pages. https://doi.org/10.1145/3711087

## 1 Introduction

In recent years, non-fungible tokens (NFTs) have emerged as a novel mechanism for digital asset ownership and exchange [123], fueling the growth of platforms like OpenSea [129]. While economic factors such as cryptocurrency prices and monetary policies fundamentally shape NFT market activity and popularity [17], the success of NFT projects also depends heavily on trust-based interactions within their communities [32, 114]. Our research therefore examines how *trust* is constructed and maintained in NFT communities, particularly given the recent decline in NFT popularity that spans both economic and social dimensions [16, 40]. From a social computing perspective, this decline offers an opportunity to investigate how trust mechanisms influence NFT adoption and engagement [15, 93, 131]. As a cornerstone of digital economies, trust facilitates *transactions*, *interactions*, and *community evolution* [67, 85]. Understanding the dynamics of trust, especially during market fluctuations, is essential for sustainable NFT ecosystem development.

The technological architecture of NFTs—rooted in blockchain—provides built-in trust protection by guaranteeing tamper-proof ownership and verifiable transaction history. Despite their technological safeguards, NFT communities face persistent trust challenges. These communities, where stakeholders interact, transact, and collaborate to promote project value [15, 21, 34], grapple with issues that undermine trust. Price volatility [74], wash trading [128], hype-driven speculation [96], and rampant phishing attempts [134] continue to erode confidence. Amid these trust challenges and broader economic fluctuations, NFT market transaction volumes plummeted in 2023–2024 [5, 7]. These market dynamics have intensified stakeholders' concerns, casting doubt on the fundamental value proposition of NFTs and the viability of their market [55, 80, 127]. Understanding and addressing these trust issues is crucial for the future of NFT technology and its potential applications in digital economies.

Trust issues in NFT communities have attracted significant research attention. However, existing studies predominantly focus on technical and commercial aspects [50, 128], with perspectives primarily derived from Western contexts in examining the global landscape of NFT communities [23, 48, 86]. This approach frequently understates the crucial role of socio-cultural factors in shaping psychological state of trust [92, 103, 138]. Given the variability of trust formation across cultures [76, 92], a culturally specific investigation from a humanistic perspective is essential, yet absent.

We address this knowledge gap by examining how socio-cultural factors shape trust dynamics in NFT communities, focusing on the Chinese context. NFTs are globally traded digital assets, yet Chinese NFT communities have emerged as a distinct segment of the global NFT ecosystem, showing steady growth and distinctive participation patterns [6, 20, 64, 65]. These communities exhibit unique dynamics in trust-building and social interactions that reflect traditional Chinese cultural values, particularly *Confucian virtues* [77, 133]. Confucian philosophy emphasizes two key concepts for trust-building: *guanxi* (a spectrum of interpersonal relationships) [118] and *mianzi* (social status earned through accomplishments or given as respect) [122]. These concepts profoundly influence daily interpersonal interactions in Chinese society [39, 78]. The Chinese socio-cultural landscape differs considerably from Western contexts [19, 22, 87]. Our exploration of these unique factors aims to find insights into trust dynamics that extant literature often overlooks [29, 73, 103].

Two research questions serve as our guides in this paper's investigation:

- *RQ1:* How do socio-cultural factors (e.g., *Confucianism*, *guanxi*, and *mianzi*) influence trust perceptions within Chinese NFT communities?
- *RQ2:* What challenges do people face in building and maintaining trust within Chinese NFT communities?

To get critical insights, we adopted a dual-pronged strategy. First, we conducted a content analysis of discussions on WeChat, a platform commonly used by Chinese NFT stakeholders [131], to gain a broad view of trust dynamics. Second, we conducted semi-structured interviews with 21 NFT stakeholders to obtain detailed insights. Our analysis identified three dimensions of trust—*technological*, *institutional*, and *social*—each affected by unique challenges and deeply influenced by the *Five Confucian Virtues*: *benevolence*, *righteousness*, *propriety*, *wisdom*, and *integrity* [77, 133].

Our research reveals a critical insight into Chinese NFT communities: Despite regulatory constraints and a rapidly evolving tech-enabled market, community members prioritize personal interactions and traditional moral principles for trust-building. This finding stresses the enduring influence of socio-cultural norms in the digital economy. Deeply cultural Confucian virtues continue to shape trust dynamics within these digital spaces. Specifically, the cultural constructs of *guanxi* and *mianzi* play pivotal roles. These traditional concepts retain their significance, even in the context of advanced digital asset markets. While these socio-cultural norms have facilitated the growth and cohesion of Chinese NFT communities, they have also presented a risk of trust-related crises. Excessive dependence on social ties and group decision-making poses risks to NFT communities. It can stifle individual critical thinking and foster "groupthink." These conditions create echo chambers within the community. In such environments, misinformation and deception spread rapidly. Consequently, trust within the community erodes, undermining its stability and value.

To mitigate these risks and reduce the potential for community disintegration, we propose a set of guidelines aimed at fostering trust within Chinese NFT communities. These guidelines could potentially be invaluable for developers, designers, and community organizers seeking to navigate the unique socio-cultural landscape of China's NFT market. Our study identifies trust issues specific to China's NFT context. However, these challenges are likely to have parallels in global NFT communities. This similarity suggests broader applicability for our trust management strategies. Our findings could offer valuable insights and advantages to NFT communities worldwide. Thus, our research contributes not only to understanding Chinese NFT dynamics but also to improving trust management in the international NFT domain.

Our research makes three main contributions to the field of CSCW and social computing:

(1) We shed light on the socio-cultural influences on trust perceptions within Chinese NFT communities. This insight enriches existing literature by providing a novel cultural perspective.
(2) We examine trust-building strategies rooted in Confucian values. Our analysis reveals both the formation mechanisms and the potential vulnerabilities of trust within these communities.
(3) We develop guidelines and design implications for trust building. These recommendations target Chinese NFT communities specifically but hold potential for global NFT and other online communities.

As we combine Eastern cultural insights with Western NFT research, our work helps create better trust management in digital economies worldwide. This approach improves our understanding of trust and gives researchers and practitioners tools to handle global digital communities.

## 2 Background and Related Works

Given that the topic of trust in Chinese NFT communities is interdisciplinary, we present relevant research from three perspectives: socio-economic, socio-technical, and socio-cultural. We have organized the extant literature into three respective subsections that mirror these perspectives: 1) *NFT communities and stakeholders*; 2) *trust issues in blockchain ecosystem*; and 3) *trust and guanxi in Chinese society*. Within each subsection, we explore relevant concepts from the macro to micro levels and conclude with a discussion on the identified research gaps and our research motivations. This structure offers a comprehensive preparation for understanding how socio-cultural factors influence the trust dynamics within Chinese NFT communities.

### 2.1 NFT Communities and Stakeholders

*NFTs.* A non-fungible token (NFT) is described as "*a cryptographically unique, indivisible, irreplaceable, and verifiable token that represents a given asset, whether digital or physical, on a blockchain*" [123]. NFTs first garnered widespread attention through the blockchain game CryptoKitties [110], which introduced the Ethereum Request for Comments (ERC) 721 standard. This standard allows each token to be individually identified using attributes such as tokenID, which facilitates the proof of ownership in the digital space [1]. NFTs have since become an integral part of the Web 3.0 digital economy [57]. Built on blockchain technology, Web 3.0—also referred to as the Semantic Web or the Decentralized Web—has enabled the emergence of new economic models like cryptocurrencies, token-based economies, and decentralized finance (DeFi) [24, 102]. In this ecosystem, NFTs act as cryptographic certificates that confirm the uniqueness and authenticity of digital assets [50].

*NFT Marketplaces.* The unique technological features of NFTs have spurred the development of a new digital economic ecosystem [57, 74], leading to the creation of various NFT trading platforms like OpenSea [9], SuperRare [10], and Blur [2]. On these platforms, participants require a personal cryptocurrency wallet, such as MetaMask [3], to engage in transactions. Creators upload and verify digital content on NFT platforms, where smart contracts mint the files as NFTs, ensuring a secure and decentralized process [27, 41]. Once validated, these unique tokens are transferred to collectors' wallets [126]. NFT marketplaces cater to diverse products, including virtual reality items [23], gaming props [93], music [123], and art [48]. Notably, personal profile picture (PFP) NFTs have carved out a unique niche. They serve as social connectors, facilitating dialogue and interaction among stakeholders, thereby fostering the growth of NFT communities [32].

*NFT Communities.* NFT communities are the primary online hubs for stakeholders to interact, transact, and collaborate, and members frequently use Web 2.0 platforms like X, Discord, and WeChat for communication [93, 114]. These communities typically form around specific projects and thrive on collective collaboration, a dynamic that has garnered the attention of researchers in the CSCW field [131]. Figure 1 illustrates the digital platforms used by members of the international (left) and Chinese (right) NFT communities for the Goblin-town project. Existing studies primarily explore two key aspects of NFT communities: who the stakeholders are [21, 130] and how they collaborate [14, 93, 131]. For instance, Almeda et al. [14] investigated how professional artists and enthusiastic collectors in NFT-based art communities collaborate to co-create artworks and enhance the community's reputation.

*Stakeholders and Community Members.* Although no complete consensus has yet been formed on the market roles of NFT stakeholders, extant literature has identified the primary member types within NFT communities as creators/launchers, brokers/moderators, and collectors/regular members [114, 131]. Launchers are responsible for initiating and establishing the community [130], moderators manage the community's daily operations [93], and regular members primarily complete tasks and participate in events organized by moderators [116]. Previous studies have demonstrated that establishing trust among members is crucial for the sustainable development of NFT communities [14, 72]. This trust facilitates seamless collaboration among the three pivotal roles [21, 114], thereby ensuring the effective functioning of these communities [25, 136].

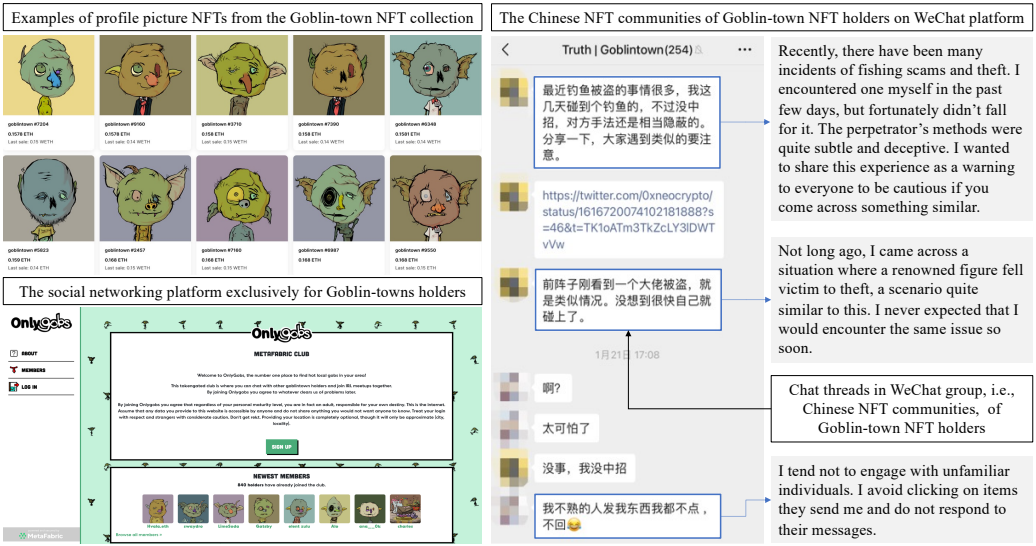

Fig. 1. Goblin-town is an example of a *personal profile picture (PFP)* NFT collection project. The top-left snapshot from OpenSea, a leading NFT marketplace, showcases a selection of PFP Goblin-town NFTs. These NFTs have a unified aesthetic with minor differences, facilitating social media recognition among holders. The bottom-left screenshot is from the official website of the Goblin-town NFT project, illustrating a social networking platform built by the launchers for all holders. The right snapshot portrays an interaction within the Goblin-town Chinese community on WeChat, where a member shares information to warn against phishing scams. Notably, WeChat is the primary platform for most Chinese NFT communities.

**Reflection.** A limitation of current research on trust dynamics within NFT communities is that it predominantly focuses on global or Western perspectives [23, 48, 86] or concentrates on a single segment, such as collectors [116] or Discord leaders [72]. In response, our study aims to examine trust dynamics within NFT communities through a non-Western lens, specifically incorporating perspectives from diverse community members. This approach seeks to enrich our understanding of the trust-building processes in NFT communities more thoroughly.

### 2.2 Trust Issues in Blockchain Ecosystem

*Trust Research in CSCW.* Trust is primarily defined in existing CSCW literature as a willingness to make oneself vulnerable based on the expected actions of another party [83, 138]. This concept also implies reliance on a dependable partner and belief in a party's integrity, benevolence, capability, and predictability [53, 88]. With the growth of the global digital economy, online communities are increasingly facing scandals that erode trust among participants [84, 124]. The CSCW community has been actively investigating decentralized web (DWeb) and blockchain technologies as alternatives that can mitigate increasing distrust in existing digital platforms [46, 49, 105, 106, 109]. For instance, Walker et al. [124] have explored the potential for digital publication through DWeb-based communal structures. Despite their potential, blockchain applications face complex trust challenges that have stirred considerable debate within the CSCW community.

*Trust Paradox in Blockchain Ecosystem.* A particular paradox exists in the technological essence of the decentralized blockchain architecture [43], which has proven to be a double-edged sword for developing the NFT space [15, 110]. While blockchain technology intends to foster a "trustless" economic environment through its immutable, transparent, and traceable digital ledger methods [51, 124], certain features can also expose the ecosystem to risks [131]. In theory, the term "trustless" suggests a shift from reliance on interpersonal or institutional trust to the integrity of blockchain code and algorithms [50, 109]. Nevertheless, in practice, such models give a paradoxical rise to new issues like phishing sites [134], counterfeit tokens [25, 42], and market manipulation [34], which collectively lead to a severe crisis of trust within the macro blockchain ecosystem and the micro NFT communities [44, 59, 106].

*Trust Issues in NFT Communities.* The trust issues in NFT communities have garnered attention from CSCW researchers, whose current explorations mainly fall into two areas: *identifying trust-related challenges in NFT communities* [86, 114] and *proposing strategies to foster trust* [37, 95, 137]. For instance, Sas et al. [105] pioneered applying trust theories to blockchain applications, creating an analytical framework categorizing trust into three types: *technological*, *institutional*, and *social*. This framework paves the way for understanding the interactions among blockchain stakeholders. As NFT communities evolved, Sharma et al. [114] observed specific challenges, such as steep learning curves and prevalent scams, that eroded trust among community members. Building on these findings, Xiao et al. [131] have proposed introducing centralized regulation to mitigate the centralization-decentralization dilemma and thereby enhance trust within these communities.

**Reflection.** However, investigations into trust and security within NFT communities typically focus on technical and business perspectives, often neglecting the crucial role of socio-cultural factors in shaping trust as a psychological construct [92, 103]. The lack of a humanistic perspective has, to a certain extent, limited the efficacy of trust-building attempts by launchers and moderators [14]. Our study addresses this oversight by examining the impact of socio-cultural elements on trust, using the CSCW trust analysis framework proposed by Sas et al. [105]. Specifically, we explore trust perceptions and practices in Chinese NFT communities, where trust-building and socio-cultural factors interact in a unique way [19, 29]. These intricate interactions are elaborated on in the forthcoming subsection.

### 2.3 Trust and Guanxi in Chinese Society

*Trust in Chinese Society.* Trust is essential in personal and commercial relationships in Chinese society, shaping how individuals, organizations, and technology interact [79, 115]. Given that Chinese society is characterized by high uncertainty avoidance, people tend to communicate less information and take conservative actions for self-protection [19, 73]. Thus, in the Chinese context, trust is critical when adopting new technologies [19, 36, 87]. For instance, in their study on vlogging in China, Chen et al. [36] found that trust in the platform and the community affects the level of user participation and the sharing of personal information. Trust boosts not only the enthusiasm of Chinese individuals for engaging in online communities but also their readiness to embrace new technological tools [63, 69]. For example, Wang et al. [125] discovered that the willingness of Chinese consumers to participate in transactions within the NFT market increases significantly when they trust the value of cryptocurrency and NFTs.

*Maintaining Guanxi for Building Trust.* The emphasis on trust in Chinese society is deeply intertwined with its rich cultural heritage and is intimately related to the concept of *guanxi* [22, 35, 38, 56, 118]. *Guanxi* represents a spectrum of relationships, from close familial ties to distant acquaintances [118]. These relationships require more effort to establish and maintain as they deepen [38]. Within this network, Chinese individuals share essential resources for survival and success to accumulate *guanxi* capital [56]. This capital is crucial for building and maintaining trust, which in turn facilitates smooth business collaborations and harmonious social interactions [22].

*Guanxi Capital and Confucian Virtues.* Following *Confucian virtues* is a highly effective strategy to build *guanxi* capital [78, 79]. These virtues, which prioritize harmony, respect, and ethical behavior in interpersonal relationships, have been well documented [77, 133]. Studies have shown that in Chinese organizations, following Confucian values can improve individuals' social reputation, known as *mianzi*, and that of others, thereby strengthening *guanxi* capital and nurturing trust within social circles [29, 122]. This dynamic is especially evident in emerging organizations using new technologies, such as NFT communities [64]. In Chinese NFT communities, *guanxi* and *mianzi* are critical to building trust among launchers, moderators, and regular members, influencing the development and sustainability of Chinese NFT communities [35].

**Reflection.** Prior studies have shown that different NFT communities develop unique cultural practices that cater to particular member values, leading to the prevalence of distinct sub-networks [13, 14]. As such, research on trust in NFT communities should be based on specific socio-cultural contexts [93, 114]. However, despite the unique interplay between socio-cultural factors and trust in China, trust-building in Chinese NFT communities remains underexplored [64]. Our study aims to fill this gap by focusing on the trust dynamics within Chinese NFT communities. Understanding these dynamics is vital not only for stakeholders in China but also for global NFT launchers and moderators who engage with Chinese community members [121, 125]. Thus, the insights derived from our study have the potential to inform the development of NFT communities on a global scale and to enhance other types of online communities.

## 3 Method

This research aims to enhance the understanding of trust formation within Chinese NFT communities through a dual-pronged approach: content analysis (CA) and semi-structured interviews (SSIs) (see Figure 2). We seek to examine how socio-cultural factors influence trust perceptions and practices, identify the challenges in building trust, and propose contextually relevant guidance to nurture and sustain trust within Chinese NFT communities. Given the sensitivity of our research topic, we obtained approval from our university's Institutional Review Board (IRB) before any data collection and analysis procedures to comply with standard ethical guidelines. The following subsections detail our research methods and the analysis procedures.

### 3.1 Content Analysis (CA)

We conducted the initial conceptual investigation using CA due to its efficacy in condensing viewpoints from a wide range of stakeholders that otherwise would be unexplored [91, 117]. Moreover, CA is a non-reactive research technique [68] that allows us to capture members' authentic sentiments unobtrusively [60]. In the CA study, *a priori* codes are generated [70, 120], thereby providing a foundation for developing an effective interview protocol for our subsequent semi-structured interviews. We next detail our data collection, data description, and analysis pipeline.

*3.1.1 Data Collection.* For the CA, we exclusively collected chat threads from Chinese NFT communities on WeChat, favoring its strong localized engagement over more globally oriented platforms like Discord and X. Language barriers are the main factor making WeChat a more active hub for Chinese members. To enhance the representativeness of our dataset, we adapted the methodology from Kim et al. [72] to select Chinese NFT communities with diverse trust levels. We assessed trust using four indicators: First, we examined the longevity of NFT projects, as this serves as a crucial measure of reliability, particularly important in markets known for "rug pulls", where projects are abruptly abandoned [112]. Second, we evaluated community engagement, gauged by the number of Chinese members and the intensity of their interactions. This indicator is significant, as higher engagement levels often suggest greater trust [137]. Third, we examined transaction records, noting that higher trading volumes are typically associated with elevated trust [58]. Lastly, we measured the community's perception of trust through surveys conducted with moderators and members, categorizing communities into different trust tiers based on the most commonly reported levels.

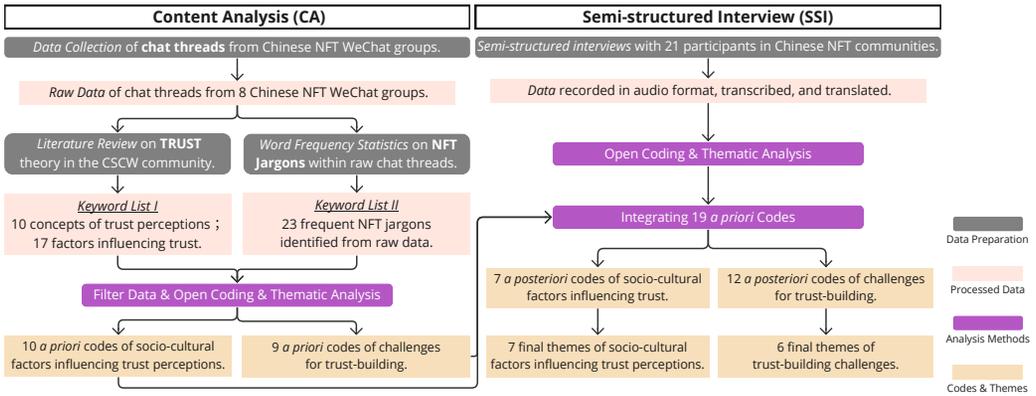

Fig. 2. Two-Fold Analysis Workflow: 1) Content Analysis (CA): Review of trust theory within social computing and text analysis from eight WeChat groups, leading to the identification of *a priori* codes for trust perception and trust-building challenges. 2) Semi-structured Interviews (SSIs): Engagement with 21 Chinese NFT community participants using *a priori* codes to conduct interviews. The thematic analysis produced *a posteriori* codes, enhancing the understanding of trust perceptions and associated challenges.

We applied the indicators and selected eight Chinese WeChat groups as our data sources. Among these, we included three communities that demonstrate high levels of trust. These communities are characterized by dedicated members who have remained active contributors even during market downturns, as the DegenToonz Chinese community exemplifies.[1] We also included three communities with moderate levels of trust, where the influence of project launchers waned over time, leading to a gradual decline in member engagement. This was observed in the Mimic Shhans community.[2] Additionally, we considered two communities that exhibit low levels of trust. These communities are notable for their complete autonomy due to unresolved conflicts between project launchers and community members. The Nyolings community provides an observed case of this dynamic.[3]

The first author joined the eight Chinese WeChat groups to collect data, ensuring that ethical research practices were adhered to. We secured ethical approval before collecting the data, and informed group administrators about the research to gain their support. We then distributed consent forms to the group members, which explained the study and the participants' rights. Only members who gave informed consent contributed data. To ensure confidentiality, we anonymized all personal information. Then, the first author used the data management application 3uTools to gather chat threads of the members' discussions.[4] As a result, our WeChat dataset accurately captures the unique trust perceptions and challenges within Chinese NFT communities. The details of these WeChat groups are presented in Table 1.

---

[1]https://degentoonz.io/
[2]https://www.mimicshhans.com/
[3]https://www.nyolings.io/
[4]http://www.3u.com/

Table 1. The table summarizes data from eight sampled Chinese NFT communities on WeChat, including each project's name, mint date, the size and duration of its community, and the number of chat threads before and after filtering with two keyword lists. It also includes the total trading volumes of the NFT projects at the time of data sampling and the trust levels as self-reported by community members.

| NFT Project Name | No. of Raw Chat Threads | No. of Textual Chat Threads | No. of Filtered Chat Threads | WeChat Group Size | Total Volume (ETH) | Mint Date | Sampling Start Date | Sampling End Date | Self-reported Trust Level |
|---|---|---|---|---|---|---|---|---|---|
| Soulda | 65500 | 39745 | 5122 | 451 | 933 | Jul-22 | 2022/9/28 | 2023/1/10 | Moderate |
| Mfers | 65500 | 47646 | 4636 | 363 | 72, 305 | Nov-21 | 2022/7/20 | 2023/4/14 | High |
| Goblintown | 16668 | 11404 | 1467 | 256 | 64,757 | May-22 | 2022/12/5 | 2023/4/30 | High |
| BoxCatPlanet | 32905 | 22568 | 1896 | 247 | 346 | Sep-22 | 2022/11/6 | 2023/4/30 | Low |
| DegenToonz | 65500 | 46424 | 3992 | 399 | 25,847 | Aug-22 | 2022/11/7 | 2023/1/6 | High |
| Mimic Shhans | 65500 | 41573 | 3917 | 466 | 430 | Oct-22 | 2022/7/11 | 2022/9/28 | Moderate |
| Nyolings | 31057 | 22273 | 2584 | 198 | 2870 | Oct-22 | 2022/11/27 | 2023/4/29 | Low |
| Ape Reunion | 26040 | 17098 | 1921 | 282 | 3,113 | Apr-22 | 2022/10/15 | 2023/4/30 | Moderate |
| Total | 368670 | 248731 | 25535 | 2662 | N/A | N/A | N/A | N/A | N/A |

*3.1.2 Data Description.* We collected raw data from chat threads of eight Wechat groups from July 11, 2022, to April 30, 2023, which was chronologically arranged in Excel spreadsheets. The average size of the NFT communities sampled is 332.75 (SD = 94.22). Each record includes the group ID, timestamp, WeChat account, and message content.

Mirroring previous studies, we implemented keyword filtering techniques to exclude irrelevant chat threads from our collection [138]. We intentionally broadened our keyword selection to avoid omitting relevant information. Specifically, we devised two lists of keywords to guide our filtering process. *Keyword list I* draws upon a survey of 21 research papers on trust concepts (e.g., responsibility, respect, safety, and comfort) analyzed via the PRISMA method [104]. These works cover trust studies on social networks, virtual communities, and the blockchain ecosystem (please refer to Table 5 in Appendix A.1), enabling us to distill associated concepts of trust perception (n = 10) and influential factors (n = 17). *Keyword list II* identifies the common *jargon* of NFT communities (e.g., FOMO, fud, rug pulls, family, and airdrop) derived from word frequency statistics on raw data of chat threads from the eight WeChat groups, locating comments directly concerning trust perceptions and challenges within NFT communities (please refer to Table 6 in Appendix A.2).

Using two predefined sets of keywords, we filtered the data to encompass 25,535 text-based conversations from 2,662 members across the eight communities (refer to Table 1). The average number of words in each chat thread in the filtered data set is 18.44 (SD = 16.04). In particular, the filtered dataset captures a wide spectrum of user engagement levels, from active members (who initiated more than 200 conversations) to lurkers (who never generated chat threads in conversations). Despite this range, most are active community members (55.91%). More precisely, 1.40%, 23.76%, and 74.84% of chat threads are generated, respectively, by launchers, moderators, and regular members.

*3.1.3 Analysis Pipeline.* This study uses thematic analysis to delve into our qualitative dataset inductively [28]. The pipeline for generating the codebook unfolds as follows: First, two of the authors, hereafter referred to as analysts, randomly selected 400 conversations (around 4,000 chat threads) from the sampling dataset to acquaint themselves with the content. Random sampling is commonly applied in qualitative studies, particularly when dealing with voluminous data [81, 101]. Second, the analysts immersed themselves in the original, unsampled dataset to develop their initial codes independently. During the separate open-coding process, the analysts generated codes bearing questions such as "*Why does this person feel trust or distrust towards other stakeholders?*"

Table 2. This table summarizes the roles of launchers, moderators, and regular members within NFT communities and illustrates how their practices contribute to trust-building.

| Community Members | Roles Within NFT Communities | Significance in Trust-building |
|---|---|---|
| 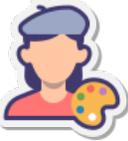 **Launchers** | • Initiate and establish NFT communities.<br>• Assemble the founding team.<br>• Attract investment and funding.<br>• Develop management frameworks post-launch. | Responsible launchers are crucial for instilling trust among members in the community's development. This trust remains strong even during market downturns, helping members maintain a positive long-term outlook on the NFT ecosystem. |
| 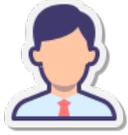 **Moderators** | • Manage daily operations of the community.<br>• Act as intermediaries between launchers and regular members.<br>• Disseminate updates on NFT projects via relevant platforms.<br>• Organize community events. | Competent moderators are crucial for reinforcing trust within NFT communities through effective daily management. They maintain this trust by organizing online activities and responding to the needs and suggestions of regular members. |
| 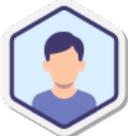 **Regular Members** | • Participate in activities organized by moderators.<br>• Have flexible involvement with no fixed responsibilities or special privileges.<br>• Propose development ideas to moderators and launchers.<br>• Create derivatives with their NFTs. | Active members play a vital role in building and maintaining interpersonal trust. Through successful collaborations and friendly, sincere communication, they enhance mutual trust. |

and "*What are the hidden cultural factors making this person have such feelings?*" The analysts then applied a qualitative analysis tool, ATLAS.ti 23, to manage and organize their findings.[5]

The whole team discussed the findings during biweekly group meetings. All authors participated in classifying the codes using the affinity diagram [62] technique through Google spreadsheets and Miro.[6] During this process, one domain expert from NFT marketplaces was consulted to aid our understanding of relevant jargon and terminology. After two months of iterative optimization of the initial codes through regular meetings, the analysts collectively distilled the codes into coherent themes. We used the *Cohen's Kappa* metric [26] to demonstrate the final thematic outcomes' inter-rater reliability (IRR) [18]. A Kappa value of 0.92 signifies almost perfect agreement among all team members regarding the *a priori* codes. The high Kappa value is due to the two analysts being Chinese native speakers with a similar understanding of the socio-cultural factors that influence trust in Chinese society.

### 3.2 Semi-Structured Interviews (SSIs)

By incorporating insights from previous research on NFT communities [14, 72, 114] with our CA results, we identified the significance of the behaviors and interactions of launchers, moderators, and regular members in fostering trust within Chinese NFT communities (refer to Table 2). We then conducted SSIs focusing on these three roles to investigate the impact of socio-cultural factors on trust perceptions and to unpack their challenges in building and sustaining trust. To develop our interview protocol we used the *a priori* codes, which served as foundational seed words for the interview questions. In the SSI step, we generated *a posteriori* codes as final themes by adding missed-out themes for the *a priori* codes to alleviate the limitations of the CA [70, 120]. The following subsections elaborate on the recruitment, participants' information, interview procedures, and analysis pipeline.

---
[5]https://atlasti.com/
[6]https://miro.com/app/dashboard/

*3.2.1 Recruitment.* To recruit participants, we adopted a dual-pronged approach, leveraging both *convenience* and *snowball sampling* strategies [45] to assemble a representative sample of stakeholders within Chinese NFT communities. We initiated the recruitment process by directly inviting individuals from the authors' social and professional networks who participated in Chinese NFT communities. Additionally, we posted recruitment advertisements on social media platforms, such as NFT-themed WeChat groups and on-campus bulletin boards. Subsequently, a few of our initial participants (P1, P5, P6, P11, P14, and P20), who were experienced NFT collectors or community moderators, aided the recruitment process by spreading the word among their personal and business connections. We carefully screened respondents to ensure they met the following criteria: 1) were above 18 years old, 2) submitted the signed consent form, 3) were Chinese citizens, and 4) had experience of over one year within Chinese NFT communities. The recruitment concluded upon reaching saturation when new interviewees began to repeat previous experiences and perspectives without adding new information.

*3.2.2 Participants.* Our sample is composed of 21 stakeholders who were actively involved in various Chinese NFT communities across different blockchain ecosystems, such as Ethereum, BTC, and Solana. The participants (see Table 3) include 12 males and nine females, spanning an age range from 19 to 45 (Mean = 31.28, SD = 6.92). Over half of the participants (57.14%) are from first-tier cities in developed regions of China, such as Beijing, Shanghai, and Hong Kong. Each participant, as per their self-report, has been involved in Chinese NFT communities for a period ranging from 1 to 5 years (Mean = 2.21, SD = 0.99). On average, participants are involved in multiple Chinese NFT communities (Mean = 7.57, SD = 8.46), contributing to both domestic (e.g., iBox) and global (e.g., Doodles, Soulda, and DegonToonZ) NFT projects. Most participants (n = 17) are from Chinese NFT communities associated with international projects. We ensured diversity in our sample by including community launchers, moderators, and regular members, with the majority (57.14%) being regular members.

*3.2.3 Interview Procedures.* Since participants were recruited online and located across various Chinese cities, all interviews were conducted remotely through VooV Meeting.[7] A week before each scheduled interview, we sent a bilingual outline of the interview topics to potential participants via WeChat and X. We provided interview outlines to participants in advance for three reasons. First, the economic sensitivity and controversial nature of NFT markets may make potential interviewees cautious about disclosing personal experiences to strangers [64, 121]. Early distribution of outlines mitigates these concerns, clarifies research goals, and promotes wider participation. Second, informed by previous CSCW research [60], having the outlines beforehand enables participants to prepare more thoughtful responses. Trust issues within NFT communities stem from complex interdisciplinary factors [44, 136]. Our pilot studies revealed that participants with economic or technological orientations tend to provide superficial responses when asked to provide responses from a socio-cultural perspective. Thus, providing outlines in advance encourages more reflective engagement. Lastly, to avoid the influence of participants' over-preparation, we adapted or reordered our questions based on their responses during the interviews. This flexible approach ensures the collection of rich and authentic data.

---

[7]https://voovmeeting.com/

Table 3. Demographic information of interviewees. Gender information: M represents male, and F represents female. "Yr. of Part." stands for the years of participation. "No. of Part. Comm." represents the number of Chinese NFT communities in which the respondents participate

| ID | Gender | Age | Location | Yr. of Part. | Community Role | No. of Part. Comm. | Related Blockchain |
|---|---|---|---|---|---|---|---|
| P1 | M | 25-30 | HongKong | 2.5+ | Moderator | 20 | Ethereum, BTC |
| P2 | M | 25-30 | Shanghai | 2+ | Regular member | 4 | Ethereum |
| P3 | M | 25-30 | Beijing | 1+ | Regular member | 1 | Ethereum |
| P4 | M | 25-30 | Wuhan | 2+ | Regular member | 20+ | Ethereum, BTC, Solana |
| P5 | M | 30-35 | Beijing | 3+ | Regular member | 3 | Ethereum |
| P6 | F | 30-35 | Shenzhen | 5+ | Launcher | 2 | Chinese Consortium Blockchain |
| P7 | M | 18-25 | Zhejiang | 3+ | Regular member | 18 | Chinese Consortium Blockchain, Ethereum, Conflux |
| P8 | M | 18-25 | Beijing | 3.5+ | Launcher | 4 | Ethereum, BTC |
| P9 | F | 25-30 | HongKong | 3+ | Regular member | 2 | Ethereum |
| P10 | M | 30-35 | Shenzhen | 3+ | Regular member | 2 | Ethereum, Solana |
| P11 | M | 18-25 | Hangzhou | 1 | Regular member | 20+ | Chinese Consortium Blockchain |
| P12 | F | 25-30 | Shanghai | 2+ | Moderator | 2 | Ethereum |
| P13 | M | 18-25 | Hangzhou | 1 | Regular member | 3 | Chinese Consortium Blockchain |
| P14 | F | 18-25 | Chengdu | 2+ | Regular member | 2 | Ethereum |
| P15 | F | 35+ | Beijing | 1.5+ | Launcher | 2 | Ethereum |
| P16 | F | 35+ | HongKong | 3+ | Regular member | 5 | Ethereum |
| P17 | F | 35+ | Beijing | 2+ | Regular member | 30 | Ethereum, BTC, Solana |
| P18 | F | 35+ | Hangzhou | 1+ | Moderator | 6 | Ethereum |
| P19 | M | 30-35 | Beijing | 1+ | Moderator | 8+ | Ethereum |
| P20 | F | 35+ | Beijing | 2+ | Moderator | 2 | Ethereum |
| P21 | M | 35+ | Shenzhen | 2+ | Moderator | 3 | Ethereum |

Upon participants' confirmation, we collected signed consent forms through Qualtrics XM.[8] Before each interview, the first author introduced the research topics and data privacy measures, duly answered the participants' queries (if any), and obtained their consent for audio recording and transcription. Participants then completed an online survey (refer to Appendix B) about their experiences, perceptions, and expectations regarding NFT communities, as well as their demographic information. These survey responses served as references during the interviews to ensure a comprehensive understanding of each participant's background. We organized the interview questions into four themes based on the *a priori* codes [9]: 1) Perceptions concerning social trust; 2) perceptions concerning institutional trust; 3) perceptions concerning technological trust; 4) extended topics (e.g., ideal NFT communities based on trust). Participants were encouraged to provide positive and negative insights on these themes during the interviews. For instance, we prompted them to recall specific events or behaviors that could foster or undermine their social trust. All interviews were audio recorded, transcribed verbatim, and translated with the participants' consent. As a gesture of appreciation, we provided all participants with Starbucks coupons upon completing our study.

*3.2.4 Analysis Pipeline.* The pipeline for analyzing the SSI data parallels that of the CA. We employed an integrated deductive and inductive strategy, facilitating a more comprehensive thematic analysis. First, we deductively applied the *a priori* codes as initial seeds. Next, we inductively added new themes identified from the interview data. Ultimately, we distilled seven *a posteriori* codes that encapsulate the socio-cultural factors influencing trust, as well as twelve *a posteriori* codes that represent the challenges associated with building trust. Efficiency was optimized by assigning the role of leading analysts to the two authors who conducted the interviews. These two authors independently coded the data and shared their findings with other team members during weekly group meetings. This iterative process continued for over a month until we reached a high IRR level. This was also measured by a *Cohen's Kappa* score of $k$ = 0.85, indicating a near-perfect agreement.

---
[8]https://www.qualtrics.com/au/
[9]please refer to the Supplemental File for details of interview outlines

## 4 Findings

Our findings are organized into two subsections to address the research questions we initially posed. Sec. 4.1 reveals the socio-cultural factors that affect the trust perceptions of stakeholders (RQ1), while Sec. 4.2 explores challenges in building and maintaining trust in Chinese NFT communities (RQ2). Chat threads extracted from the CA and quotes from the SSIs are *italicized*. We have also attached anonymous identifiers (refer to Table 3) to each quote to ensure the privacy of our interviewees. Furthermore, the percentage values associated with each theme indicate the frequency of that theme's occurrence in the CA and SSIs.

### 4.1 Socio-Cultural Factors Influencing Trust

This subsection explores how Chinese NFT communities adopt the technological features of NFTs and investigates the interplay between these features, the socio-cultural context, and trust perceptions. We observe that the technological intricacies of NFTs, designed for trustless economic environments, are not fully embraced by Chinese NFT communities [125]. This hesitancy primarily stems from two factors. Firstly, technological features such as anonymity and immutability may conflict with traditional Chinese trust-building norms [64], which prioritize harmonious interpersonal relationships built on Confucian virtues [19, 122] (see Table 7 in the Appendix). Secondly, many members of Chinese NFT communities are crypto novices who find NFT marketplaces less inclusive and related tools less accessible than Web 2.0 tools [25]. Consequently, these members' trust perceptions are heavily shaped by socio-cultural factors, emphasizing the importance of interpersonal interactions and community relationships.

To demonstrate how socio-cultural norms of Chinese society affect trust perceptions among stakeholders in Chinese NFT communities, we employed the trust analysis framework developed by Sas et al. [105]. This framework allows us to deconstruct trust into three dimensions, considering the roles of both the trustor and trustee (see Table 4):

(1) **Technological trust** highlights the relationship between community members and the NFT ecosystem.
(2) **Institutional trust** details the interactions between community members and the launchers or moderators.
(3) **Social trust** encapsulates the interpersonal interactions among the community members.

By applying this model, we gain insights into how socio-cultural factors intersect with the technological intricacies of NFTs, influencing trust perceptions of stakeholders within these communities.

*4.1.1 Socio-Cultural Factors Influencing Technological Trust.* In WeChat groups, we observed that technological trust within Chinese NFT communities centers around two questions: how secure members feel when using crypto tools and how confident they are in the future of the NFT ecosystem. These issues are particularly acute in China due to the government's ban on cryptocurrencies, which restricts daily crypto tool usage [132]. Consequently, many community members are unfamiliar with blockchain and NFT technologies, leading to widespread insecurity when using NFT trading platforms and cryptocurrency wallets. Instead, members often rely on *endorsements* from respected senior members within their WeChat groups and conform to the majority's behavior and attitudes. This dependence on established social networks and preference for *conformity* fundamentally influences members' perceptions of technological trust within Chinese NFT communities.

Table 4. Overview of three trust categories. Text in light green suggests the feeling of trust, text in orange indicates the trustee, and text in light purple represents the attributes (from evidence/experience/expectations) of the trustee.

| Category of Trust | Sub-themes | Definitions |
| --- | --- | --- |
| 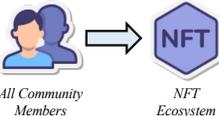 *All Community Members* → *NFT Ecosystem* **Technological Trust** | • Feeling secure in using crypto tools • Feel confident about NFT marketplaces | • Participants feel secure when using crypto tools, e.g., crypto wallets, analytics tools, validation sites, and transaction platforms. • Participants feel confident about the healthy development of the macro NFT marketplaces. |
| 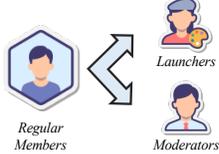 *Regular Members* ↔ *Launchers / Moderators* **Institutional Trust** | • Faith in NFT projects • Believe in launchers and moderators | • Participants have faith in the growing impact and potential appreciation of owned NFT projects. • Participants believe in the reliability and ability of launchers and moderators of NFT projects, who are expected to fulfill plans in roadmap. |
| 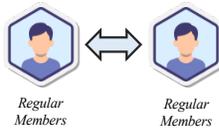 *Regular Members* ↔ *Regular Members* **Social Trust** | • Willingness for collaboration • Trust in shared information • Interpersonal closeness | • The willingness to participate in online and offline social or business activities initiated by other community members. • Participants regard the information and knowledge shared by each other as valuable and helpful. • Participants feel the closeness and intimacy between each other during interpersonal interactions. |

***Endorsement*** *(CA = 10.66%, SSI = 22.65%).* Endorsements shape perceptions of technological trust within Chinese NFT communities by helping members overcome language and knowledge barriers. In contrast to the international NFT community, whose concerns center around hacker attacks due to smart contract vulnerabilities, most WeChat group members are more concerned with their proficiency in operating crypto tools. Frequent inquiries in WeChat groups, such as "*How do I download a crypto wallet?*" and "*What translation software is most helpful?*" illustrate the basic knowledge gaps prevalent among them.

Rooted in traditional Confucian values, particularly respect for authority, members of Chinese NFT communities heavily rely on endorsements of tools and market insights from trusted senior members. As P19 opined, "*Everyone seeks authority. It is our habitual way of thinking and acting.*" In this environment, senior members serve as pivotal sources of essential knowledge. Their endorsements of crypto tools and market insights effectively enable newcomers to quickly acquire NFT basics, thereby enhancing the community members' perceptions of technological trust. For example, in WeChat groups, it is common for moderators to spontaneously alert other members with messages such as "*This link is a phishing site; please be careful,*" "*Recently, someone's NFT was stolen; do not authorize wallet permissions indiscriminately,*" or "*It is recommended to use a cold wallet for increased safety.*" As P9 stated,

> "…I heavily relied on the guidelines provided by moderators in WeChat groups when trading NFTs. They frequently recommended reliable crypto tools and alerted community members about recent crypto scams through social media. Their endorsements reassure me when using those tools…"

Prioritizing endorsements from recognized senior members over independent learning is particularly prominent within Chinese NFT communities. This trend underscores the substantial role that traditional Confucian values, such as respect for authority, play in shaping the Chinese members' trust in new technologies.

**Conformity** *(CA = 14.52%, SSI = 15.90%).* Conformity is also pivotal when perceiving technological trust in Chinese NFT communities. Our analysis reveals that although transaction records in NFT marketplaces are transparent, the anonymity of traders tends to unsettle members of Chinese communities. As P1 stated, "*There are too many scammers. When approached by unfamiliar or suspicious individuals, it is safer to ask others in the group.*" Therefore, instead of solely relying on objective data, many community members take cues from their peers, adjusting their tool selection and transaction timings accordingly. Common inquiries in WeChat groups include, "*Let me know when you all are buying NFTs*" or "*Which platform are you all currently using for NFT transactions?*" This trend suggests that technological trust within Chinese NFT communities is often the product of collective approval rather than personal understanding.

Moreover, during periods of market volatility, community members typically probe the intentions of others to inform their own decisions on whether to sell their NFT holdings. As P13 said,

> "*...My knowledge of blockchain and NFTs may be limited, but I trust the judgment of my fellow (WeChat) group members. I use the same tools and trade at the same times they do. Although they are not always right, this collective guidance gives me more confidence over going it alone...*"

As such, conforming to the practices of peers can help community members build technological trust in anonymous NFT marketplaces by fostering an exclusive *guanxi* network. Yet, these collective decisions may not always outperform those derived from transparent blockchain data.

*4.1.2 Socio-Cultural Factors Influencing Institutional Trust.* In Chinese NFT communities, institutional trust is founded on members' belief in the long-term potential of NFT projects and their respect for moderators and project launchers, who are seen as formal or informal leaders. While NFTs offer transparent provenance for digital assets in decentralized marketplaces, the lack of standardized assessment metrics for different NFT projects leads to considerable market price volatility [114, 136]. Consequently, reliable community leaders should embody the traits of a "noble person" as defined by Confucianism, providing both financial and emotional support to members. By prioritizing the overall welfare of the community (demonstrating *generosity*) and appreciating the contributions of regular members (aligning with *face culture*), these leaders greatly enhance institutional trust within the community.

**Generosity** *(CA = 19.15%, SSI = 6.41%).* Generous project launchers and moderators enhance institutional trust in Chinese NFT communities by prioritizing collective well-being. While NFTs establish ownership rights for all members, the varying token prices due to rarity and market volatility can create significant wealth disparities within the community [110]. These disparities conflict with Confucian ideals, which stress that material inequality within a group of shared interests poses a greater threat to sustainability than scarcity [77]. In WeChat groups, we have also observed that the most popular slogans among community members are "*Common Prosperity*" and "*Family Advancing and Retreating Together.*"

Generous actions by launchers and moderators—such as distributing small red envelopes (monetary gifts) in WeChat groups, organizing whitelist raffles on X, or sharing a modest portion of royalties—can foster community cohesion, bolster members' confidence in the project's long-term viability, and ultimately strengthen institutional trust. For example, P20 noted,

> "...Community members tend to trust generous launchers. For instance, DegenToonZ, an NFT project, regularly organizes coffee and dinner gatherings for the community, which have received positive responses [...] Even during bear markets, this strong foundation of trust motivates members to stand by the project..."

Generosity is crucial in building institutional trust in Chinese NFT communities. As P17 commented, these simple gestures may not always yield direct financial gains for members, but they provide a "*narrative of collective success*" that helps members remain patient and supportive as the community navigates through challenges.

***Face Culture*** (CA = 14.57%, SSI = 15.13%). In Chinese NFT communities, the practice of face culture by launchers and moderators greatly influences members' perceptions of institutional trust. Face culture, while similar to politeness, focuses on expressing recognition and appreciation for others [29, 122]. Specifically, Chinese NFT communities are often spontaneously formed, self-regulated, and decentralized online spaces that thrive on voluntary contributions from their members. While a few well-educated individuals may join official platforms like Discord or Telegram to establish global connections, most regular members work anonymously under the guidance of local moderators to support community development. These members, who have not joined official platforms, jokingly refer to themselves as "*cannon fodder*" in interviews. Their limited interaction with the core NFT project team can lead to fragile institutional trust, especially during challenging times.

Such members deeply appreciate it when launchers and moderators align with face culture by respecting their cultural identity and acknowledging their contributions. Simple gestures, such as following local linguistic norms, celebrating Chinese holidays, and using popular software in China, can help members feel respected and valued. For example, when project launchers who are native English speakers reach out to community members in Mandarin, participate in WeChat groups, or involve Chinese participants in the core council for project development decisions, they uphold face culture and build institutional trust within the community. As P1 noted,

> "...To build trust in the NFT community, project launchers must show respect to their members [...] However, it can be challenging sometimes. At least, launchers should engage with their holders on social media platforms (e.g., X), responding to comments, following each other, and avoiding a patronizing attitude..."

As such, members in Chinese NFT communities, especially those who cannot participate in official channels, greatly value being recognized and appreciated by launchers and moderators. This practice of face culture goes beyond mere politeness and is essential for enhancing and sustaining institutional trust.

*4.1.3 Socio-Cultural Factors Influencing Social Trust.* Social trust, or interpersonal trust, refers to the one-on-one trust relationships between members within NFT communities. This trust is demonstrated through three progressive levels: enthusiasm for participating in community events, trust in shared information, and the tendency for members to cooperate amicably.

While NFTs technically enhance social trust by offering transparent and immutable on-chain records, adherence to established socio-cultural norms remains a crucial factor influencing social trust perceptions within Chinese NFT communities. This dependence arises from two main factors: the cultural significance of *guanxi* in fostering interpersonal trust in Chinese society and the technical and legal controversies surrounding NFTs. For example, P1, P4, and P9 expressed doubts about investment recommendations from anonymous peers, highlighting that "*the immutability of NFT transactions prevents (me) from reversing fraudulent decisions.*"

Specifically, members in Chinese NFT communities enhance perceptions of social trust by actively building *guanxi capital*, demonstrating *ethical behaviors*, and showing *loyalty* to the project during challenging times.

**Guanxi Capital** *(CA = 10.67%, SSI = 16.67%).* In China, *guanxi* capital is an indispensable yet easily accumulated form of social capital for building trust [73]. Typically, acquiring *guanxi* capital entails following implicit social norms, such as consistently displaying small, ritualistic behaviors that reflect the community's shared identity.

Accumulating *guanxi* capital remains important even within decentralized and self-regulated Chinese NFT communities. These subtle efforts foster member interactions, encourage participation in community events and gradually strengthen community members' perceptions of social trust. For instance, accumulating *guanxi* capital can involve sending festive greetings in WeChat groups, commending community members on their contributions and achievements, maintaining a polite and tolerant attitude during disagreements, and helping others resolve their confusion or problems encountered in the NFT space. As P18 illustrated,

> "...My deepest trust within the NFT community is reserved for those members with whom I have shared the journey—collaborating on projects, navigating the highs and lows, and exchanging confidential thoughts. Our shared experiences have fostered profound social trust among us..."

Accumulating social capital among Chinese NFT community members reflects their adherence to mainstream Chinese social norms. This adherence helps alleviate the psychological pressure linked to decentralized and anonymous interactions, ultimately fostering natural empathy and a sense of belonging and enhancing social trust within the community.

**Ethical Behaviors** *(CA = 14.33%, SSI = 7.26%).* Ethical behaviors can help bolster social trust within Chinese NFT communities by mitigating the risks of misinformation and fraud. Since their inception, NFTs have been accompanied by various ethical controversies, some of which (e.g., fraud, scams, and theft) are difficult to resolve solely through automated code [47]. Additionally, certain technological features of NFTs, such as the previously discussed immutability, further complicate these ethical issues.

In response, members of Chinese NFT communities enhance social trust by practicing ethical behaviors that embody Confucian virtues. Chinese society has long revered individuals who prioritize ethical conduct over financial interests. Thus, individuals who demonstrate these ethical behaviors are seen as commendable and worthy of trust and friendship.

Specifically, engaging in ethical behaviors, such as voluntarily sharing tutorials on NFT basics, investing personal funds to create NFT derivatives, and supporting crypto scam victims, can nurture community social trust and improve individual reputations. For example, P4 shared his experience of returning a stolen NFT to its rightful owner, a kind action that garnered him widespread trust from members:

> "...I purchased a stolen NFT and returned it to its rightful owner. Following the incident, I increased my communication with other community members, which led to many new friends. Although we have never met in person, they are willing to entrust me with their private keys to assist them in minting NFTs..."

**Loyalty** *(CA = 16.10%, SSI = 15.98%).* Loyalty is crucial in Chinese NFT communities for building social trust by fostering effective collaboration among members. Chinese culture, which values dialectical reasoning and collectivism, has always revered loyalty [122]. This trait signifies patience for development and a commitment to the collective good.

Nonetheless, the underdeveloped regulatory environment for decentralized NFT marketplaces remains vulnerable to exploitation by speculators [131] who often migrate among different Chinese NFT communities when market prices fluctuate [30]. Their behavior undermines the community's sustainability and weakens social bonds among members. P10 expressed frustration with speculators, saying, "*They are selfish and disruptive, and often cause unrest within the community.*" Therefore, in Chinese NFT communities, loyal members who support the community, particularly during challenging times, are more likely to earn their peers' trust and encourage them to overcome difficulties together. As P12 noted,

> "*...I believe in loyal members. They deeply value Soulda's aesthetics and ideology, and their dedication is why I continue volunteering as a moderator for the (Soulda) Chinese community. Although they might be few, their commitment truly forms the bedrock of our community's growth...*"

### 4.2 Challenges in Building and Preserving Trust

Each dimension of trust within Chinese NFT communities faces distinct challenges (see Figure 3). Specifically, challenges to technological trust primarily stem from the technology's limitations and the immaturity of marketplaces. Institutional trust challenges are mainly due to the improper behavior and attitudes of project launchers or moderators. Meanwhile, social trust is more vulnerable to threats from internal community conflicts. The following subsections will delve into the specific challenges for each trust dimension.

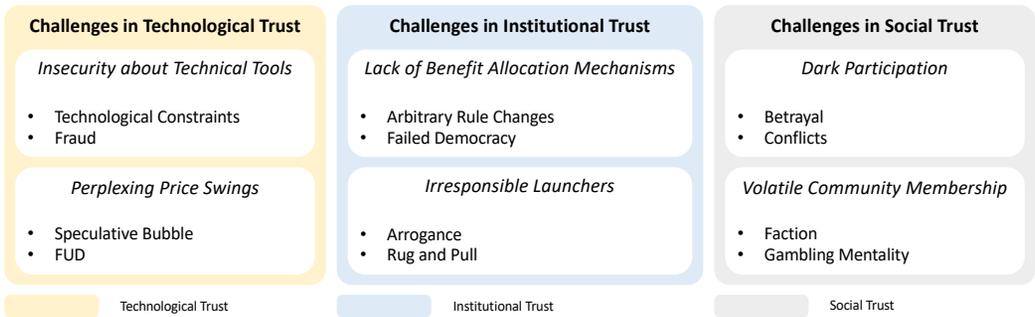

Fig. 3. Three categories of challenges to trust in Chinese NFT communities: Technological trust is affected by security concerns and unpredictable pricing. Institutional trust suffers from unclear benefit allocation and irresponsible launchers. Social trust is weakened by dark participation and volatile memberships.

*4.2.1 Challenges in Technological Trust.* Interpersonal relationships can help members within Chinese NFT communities with their initial onboarding and developing a basic trust in the technology, but they are not a panacea. The intrinsic limitations of technology and the developing nature of marketplaces introduce two main challenges to fostering strong technological trust: 1) *insecurity about technical tools* and 2) *perplexing price swings*.

These issues are prevalent in the NFT ecosystem but have a particularly significant impact in China, where the community heavily relies on interpersonal networks. Dishonest individuals exploit this reliance to manipulate public opinion and commit fraud. This exploitation amplifies panic among members who are not well-versed in NFT technology and market dynamics, further exacerbating the challenges to building technological trust.

***Insecurity About Technical Tools*** (CA = 25.44 %, SSI = 10.34%). Participants reported feeling insecure when using NFT-related tools, particularly regarding user experience and the protection of user rights. In WeChat groups, while discussions occasionally arise about transaction speed, smart contract audits, and NFT compatibility, the primary concern remains the security of crypto wallets and NFT trading platforms. This insecurity hampers the development of technological trust within Chinese NFT communities. Two primary factors contribute to this challenge: *technological constraints* and the risk of *fraud*.

Current crypto tools are limited in functionality, user interface (UI), and interaction design, failing to fully leverage the advantages of blockchain technology. As P2, a full-time product manager, commented, "*Many Web 3.0 tools have very limited application scenarios, and their interaction flows are not well-developed.*" Even worse, the principle of "*code is law*" [108] allows scammers and speculators to take advantage of vulnerabilities in smart contracts for malicious gain. In WeChat groups, people often express concerns about data manipulation, the lack of screening for phishing websites, and their inability to recover stolen NFTs. As P19 stated,

> "...*While blockchain's transparency is noteworthy, it is not completely reliable. Inscriptions in Ether or BTC may discourage malicious behavior but cannot ensure total data security or authenticity. Despite existing safeguards, there is still the potential for launchers to manipulate the data as they wish...*"

Secondly, inconsistent policies and strategies across various NFT projects and trading platforms often confuse community members. For example, they have discovered that some platforms, such as Blur [2], "*even allow stolen NFTs to be listed*", which increases the risk of fraud. Given these circumstances, community members must exercise exceptional caution with all minting-related links due to the irreversible nature of NFT transactions. This is especially critical when authorizing private keys, as any errors in this process can lead to the complete forfeiture of wallet assets. P1 succinctly captures this sentiment:

> "...*My distrust in crypto tools is primarily due to the irreversible and considerable risk of NFT theft [...] Reporting such incidents can only limit the liquidity of the stolen NFTs without providing any compensation, thereby intensifying my concerns...*"

Due to their insecurity toward crypto tools, many members heavily rely on their interpersonal networks for reliable insider information. Nevertheless, this excessive dependence may compromise technological trust. Overreliance on these networks can lead to public opinion manipulation, which, as discussed below, can diminish members' confidence in the NFT ecosystem.

***Perplexing Price Swings*** (CA = 19.30%, SSI = 20.20%). Price swings can trigger intense emotional responses among community members and erode stakeholders' confidence in Chinese NFT communities. Specifically, during market highs, stakeholders can easily be swept up in the fervor of *fear of missing out (FOMO)*, leading to speculative bubbles in NFT marketplaces [127]. As P17 stated, "*People tend to lose rationality during price booms. Without understanding the underlying blockchain technology or cultural and utilitarian value of NFTs, their strategy can become highly speculative [...] They are prone to purchase NFTs impulsively during a bull market, setting the stage for community disputes when prices fluctuate.*"

The situation takes a sharp turn when the value of the high-cost NFTs they have invested in plummets. Community members who lack adequate financial resources may quickly succumb to *fear, uncertainty, and doubt (FUD)*. This sentiment often spills onto social media platforms, triggering a negative feedback loop that undermines overall market trust, a phenomenon captured eloquently by P21:

> "...Trust within Chinese NFT communities is a delicate balance, easily disrupted by FUD [...] Negative project chatter on social media leaves members panicked about price volatility and adverse market sentiment, severely testing their trust..."

Turbulent public opinion and NFT price fluctuations often occur simultaneously and can profoundly impact technological trust within Chinese NFT communities. Due to the homogeneity in trading behaviors created by feelings of FOMO, members are more likely to empathize with one another. This intense collective emotional volatility can lead to widespread distrust in technology and the market.

*4.2.2 Challenges in Institutional Trust.* Unlike Western NFT communities, which are often driven by entertainment and consumerism [23, 114, 137], members of Chinese NFT communities emphasize economic gains and the emotional value of being respected. Therefore, the core NFT project team must ensure monetary benefits and emotional support to sustain a lasting partnership with the Chinese NFT community. However, effectively addressing these dual motivations is challenging. Two key factors primarily threaten the institutional trust of regular members. Foremost is *the absence of a robust benefit allocation mechanism*, leading to substantial disagreements among launchers, moderators, and regular members during collaborative endeavors. The second factor is *irresponsible launchers*. When project launchers disregard face culture and fail to follow through on their roadmap plans, they impair the institutional trust in Chinese NFT communities.

**Lack of Benefit Allocation Mechanisms** (CA = 18.42%, SSI = 21.18%). The absence of a clear benefit allocation mechanism reflects a lack of generosity on the launchers' part and frequently sows seeds of institutional distrust. Many Chinese NFT communities organically spring up from collectors themselves, often with local moderators voluntarily serving without compensation. In light of this, community members expect to attain long-term economic benefits through the community's sustainable development. Consequently, when launchers *arbitrarily amend profit-sharing rules* to accrue personal benefits, it can spark serious internal conflicts. As P9 observed, "*Trust towards an NFT project weakens when launchers unpredictably change plans, especially without a clear roadmap or transparent smart contract benefit rules. In severe situations, launchers may exploit community enthusiasm to break copyright-sharing promises after gaining market influence, leading to outright community collapse.*"

In more extreme scenarios, arbitrary rule changes can result in *a total failure of democracy*. This could trigger a trust implosion for NFT projects that heavily rely on community-driven efforts. These projects are primarily sustained through community self-governance and economic crowd-funding by their members. As P5 highlighted,

> "...Inadequate collective decision-making can erode trust. Key issues such as royalty rates, account sharing, project evolution, and treasury management demand community dialogue or advance notice. Furthermore, communities should align voting weights with users' NFT holdings, but some launchers disregard this principle, even manipulating votes—an unacceptable practice..."

Given the large size of Chinese NFT communities, typically comprising four to five hundred members, overly relying on interpersonal relationships for collaboration can lead to an unfair distribution of benefits. Without robust benefit allocation mechanisms, this unfairness can cause members to feel deceived or exploited, ultimately weakening institutional trust.

**Irresponsible Launchers** (CA = 14.91%, SSI = 18.72%). There are two major indicators of irresponsible project launchers within Chinese NFT communities. The first is an attitude of *arrogance*, and the second is engaging in the deceptive practices known as *rug pulls*.

Proc. ACM Hum.-Comput. Interact., Vol. 9, No. 2, Article CSCW189. Publication date: April 2025.

Arrogance is evaluated by whether project launchers recognize the *mianzi* (social status earned through accomplishments or granted as a sign of respect) of community members, particularly veterans and key contributors, in their communications. For instance, P18 described the profound harm she experienced when the launchers of a project allowed rumors against her: "*Unsupportive launchers during attacks or rumors can greatly harm trust. As an unpaid moderator, I encountered libel on social media. Despite seeking help from the launchers, they dismissed my concerns, failed to investigate, and subsequently removed me from my role. This experience killed my enthusiasm for NFT community involvement.*"

Another less severe case of arrogance arises when launchers are reluctant to learn from and integrate with community members, especially those who are newcomers to the NFT market, despite their success in other domains. Without prompt correction, this attitude can lead to further ill-conceived plans, such as over-promotion of the project or mismanaged expectations, which can trigger a breakdown in institutional trust. As illustrated by P12, "*Blind imitation of success in other industries does not ensure trust. While learning from other domains is worthwhile, launchers must thoroughly grasp NFTs and smart contracts' technology and rules. This understanding helps establish promising project paths.*"

Judging a rug pull is more straightforward: this term refers to the launchers' failure to execute the plans detailed in the roadmap, which is crucial for the community's sustainable development. When a rug pull happens, launchers often abruptly terminate the NFT project after reaping profits, stranding all community members without support or recourse. As P19 commented,

> "*...The rug pull tactic often entails rapidly deploying all marketing strategies (before launch), like releasing a roadmap or engaging with users on X spaces. However, the real damage occurs post-launch, when these launchers pocket the funds and entirely abandon the project and community, severely damaging our trust...*"

*4.2.3 Challenges in Social Trust.* Building social trust within Chinese NFT communities faces two primary challenges, both related to financial incentives. The first, arguably the more challenging, is *dark participation*. This refers to specific community members who, in their pursuit of personal gain, incite conflicts within the community, destabilizing it. The second factor is speculators, whose sole focus is monetary profit. These individuals are relatively easier to identify and manage as they have *volatile community membership*, leaving when the market turns downward.

**Dark Participation of Community Members** *(CA = 14.91%, SSI = 10.84%).* While Chinese NFT communities greatly respect the Confucian principles of benevolence and righteousness and value harmony, adherence to these socio-cultural norms is not universal among community members. There are individuals who engage in dark participation, characterized by acts of *betrayal* and irreconcilable internal *conflicts*.

Betrayal happens when community members prioritize personal gain over collective interests, resorting to deception and aggression against their peers and damaging social trust. For example, some may commit fraud using fictitious offline identities, coerce community members into disadvantageous transactions, or disseminate harmful rumors. As pointed out by P11, "*Betrayal greatly erodes social trust. Some members exploit information gaps and others' trust, issuing fake news to mislead trading decisions and profit from the resulting NFT price volatility.*"

Furthermore, irreconcilable internal conflicts over project development and resource usage, though less damaging than outright betrayal, can also diminish social trust. If left unaddressed by project launchers or moderators, these conflicts can escalate into a full-blown trust crisis. P2 noted,

> "...Interest conflicts and emotional disputes can stem from promoting rival NFT projects, disagreements on development plans, or respect issues between new and established members. These disputes, escalating on platforms like WeChat and X, can trigger a community trust crisis..."

In Chinese society, where a strong emphasis is placed on insider support and assistance, community members may be particularly vulnerable to these malicious acts. When conflicts or betrayal occur, it profoundly impacts both the individuals affected and the community at large.

**Volatile Community Membership** *(CA = 7.02%, SSI = 18.72%).* In Chinese NFT communities, volatile community membership significantly undermines social trust. Frequent arrivals and departures of members are common, yet it is the departures that cause the most damage to the level of social trust. Some members are forced out by internal *factions*, whereas individuals with a *gambling mentality* tend to voluntarily exit when their financial expectations are not met.

For instance, some members might secretly form exclusive factions without notifying others or the community moderators. This might not pose an issue in the absence of conflicting interests. However, disruptive actions may occur when these factions' interests clash with those of other community members or the wider NFT community. They might initiate divisive activities or exclude specific members, pressuring them to exit the community. As reported by P9, "*Destructive sub-groups can disrupt trust. Even united, harmonious Chinese NFT communities are vulnerable to factions that isolate members, create discord, and manipulate public sentiment for personal gain. The truth, often concealed by these factions, makes resolving such issues challenging.*"

Another scenario involves members with a gambling mentality who prioritize economic gains over cultural, aesthetic, or social considerations. If these individuals do not realize their anticipated financial gains, they might withdraw from the NFT community. For example, P11 and P13 openly admitted that their main motivation for joining NFT communities was to make money, and if they failed, they would leave. They demonstrate a lack of patience for the long-term development of the NFT community, stating that "*trading NFTs is just like sneaker flipping [...] holding on too long will only result in heavy losses.*" Frequent withdrawals of this nature can heighten the insecurity linked with anonymous social interactions and severely tarnish interpersonal trust among all community members. As P15 mentioned,

> "...Social trust is challenged in speculator-dominated communities. Trust-building, a long-term process, is often undermined by speculators' impatience for project development and minimal community contribution. Focused only on price fluctuations and lacking project loyalty, they pose a severe operational challenge to the community..."

## 5 Discussion

As Kennedy et al. [71] aptly state, "*People are not lab rats, and their engagement with designed artifacts does not occur in situations free of cultural or social values and contexts.*" Previous studies have highlighted the existence of distinct sub-networks within the global NFT community, each developing specific cultural practices that align with particular contexts and social values [14, 136]. Building on this foundation, our analysis explores how socio-cultural factors influence trust among members of Chinese NFT communities.

We found that members within Chinese NFT communities place a heavy reliance on interpersonal relationships to build trust in technology, institutions, and fellow members due to language barriers, the novelty of technology, and the nascent state of marketplaces. However, social relationships are not a panacea; an over-reliance on *guanxi* introduces unique challenges in maintaining trust within Chinese NFT communities.

Based on our findings, in this section, we compare the trust challenges Chinese NFT communities face with those in the broader NFT ecosystem (Sec. 5.1). Accordingly, we propose guidelines (Sec. 5.2) and design implications (Sec. 5.3) primarily aimed at building trust within Chinese NFT communities. These guidelines focus on three key roles within NFT communities, launchers, moderators, and regular members, and are aligned with traditional Chinese norms, such as Confucian virtues. Although these strategies are tailored to a Chinese audience, they could also benefit the broader NFT ecosystem, given the overlap and relevance of trust challenges.

## 5.1 Building Trust in NFT Communities: Socio-Cultural Insights from China to the World

Here, we compare the trust challenges Chinese NFT communities face with those in other cultural contexts, underscoring the potential global significance of our trust-building insights.

*5.1.1 Unique Trust-Building Challenges in Chinese NFT Communities.* Chinese NFT communities face unique trust challenges stemming from a blend of novel technologies and a profound cultural legacy [29, 39].

One notable obstacle within Chinese NFT communities is the higher percentage of crypto novices and their idealized expectations of community leaders. With a general lack of confidence in assessing expertise, members of the Chinese NFT community emphasize the intentions and actions of their leaders. They expect thorough and patient onboarding processes, alongside demonstrations of humility, inclusivity, and a strong sense of accountability. These expectations are rooted in the Confucian principle of *guanxi*, which underscores social support and interpersonal harmony [35, 122]. In contrast, Western NFT communities often prioritize self-reliance and innovation, urging members to learn and grow independently [92]. Recognizing these cultural nuances is essential for effectively managing cross-cultural NFT communities on platforms like Discord and Telegram [14, 103]. By integrating these understandings, leaders of such communities can enhance their engagement with Chinese members and cultivate a more inclusive atmosphere.

Furthermore, the high-context nature of Chinese society [73], which values stable and close social networks, leads to an over-reliance on interpersonal relationships. This dependence can result in distinct trust challenges, particularly during periods of price volatility or instances of malicious behavior. These situations often trigger strong collective emotional responses within the community, as members tend to act collectively and exhibit high levels of empathy influenced by Confucian virtues. This collective emotional surge can exacerbate panic over market volatility and lead to widespread irrational trading, ultimately hindering trust-building within Chinese NFT communities. Moreover, given the far-reaching influence of Confucianism across East Asia, including in countries like Japan and Korea [87, 135], similar patterns may emerge in other East Asian NFT communities. Therefore, launchers and moderators in those communities should incorporate socio-cultural factors into their regulation strategies.

*5.1.2 Common Trust Challenges Across NFT Ecosystems.* Despite their unique characteristics, Chinese NFT communities share several trust-building challenges with the broader NFT ecosystem.

First, trust challenges arising from the centralized-decentralized governance dilemma are common, making it difficult to balance control and autonomy [51, 131]. This decentralized vision complicates the effective regulation of NFT communities, leading to hype [96], rug pulls [112], and fraud [107]. As a result, community members struggle to assess the credibility of information and make rational investment decisions. For instance, Xiao et al. [131] and Catlow et al. [33] have identified that popular slang in NFT communities on platforms like Discord and X often incites impulsive buying, potentially leading to economic bubbles similar to the Dutch tulip mania. This is consistent with our findings in WeChat groups, where certain members may manipulate public opinion, causing the community to experience FUD (fear, uncertainty, and doubt) or FOMO (fear of missing out), indirectly exacerbating NFT market volatility.

Second, concerns about the security of using crypto tools are shared by both Chinese NFT communities and the broader NFT ecosystem. The prevalence of phishing websites, counterfeit NFTs, and hacking attacks poses severe trust challenges. For example, Sharma et al. [113, 114] noted that while smart contracts simplify NFT transactions and reduce the need for intermediaries, issues such as wallet custody, password recovery, and disputes over stolen NFTs remain unresolved. These problems are widespread in different NFT communities and require enhancing the user experience of crypto tools and their integration with relevant Web 2.0 tools.

Third, the misconduct of community leaders, such as launchers or moderators, can undermine trust. Although members' motivations in different NFT communities may vary—some are there for consumption, entertainment, and socializing [23, 137], while others focus on investment, profit, or career opportunities [93, 124]—the ability of launchers and moderators to manage relationships with community members profoundly impacts trust. For example, Kim et al. [72] found in their study of NFT communities on Discord that leaders who exhibit arrogance, emotional instability, and irresponsibility damage trust within the community. This aligns with expectations in Chinese NFT communities, where launchers are expected to emotionally support community members to maintain trust and improve community cohesion.

## 5.2 Applying Socio-Cultural Factors of Trust: Guidelines for Trust Building in Chinese NFT Communities

We propose trust-building guidelines tailored for roles within Chinese NFT communities, grounded in socio-cultural factors that greatly shape trust perception. Given the profound impact of Confucianism on East Asian society and its partial convergence with universal principles of trust, such as kindness and authenticity [66, 138], these guidelines can benefit cross-cultural NFT communities. Furthermore, the increasing participation of Chinese individuals in the global NFT market underscores the value of our guidelines for international launchers and moderators seeking effective collaboration with Chinese members. To this end, we have designed an action framework (see Figure 4) based on the guidelines to enhance trust among the three key roles within NFT communities.

*5.2.1 Guidelines for Building Technological Trust.* We propose a unified strategy to help moderators and regular members establish technological trust from the outset. The initial and most critical step for them is to understand the fundamentals of blockchain and NFTs and become proficient in using crypto tools, such as wallets and trading platforms. Adequate knowledge could reduce the risks associated with overreliance on interpersonal relationships, which can trigger technological trust issues. Furthermore, moderators must be thoroughly informed about the long-term strategies of the NFT projects they support and reject invitations from entities with fraudulent intentions. Meanwhile, regular members must familiarize themselves with NFT scams and maintain vigilance against risks, including phishing attacks and identity theft.

At the same time, launchers must assume multifaceted responsibilities to nurture technological trust. First, they must secure resources, such as funding, technical support, and design teams, before initiating NFT projects. Second, NFT launchers must understand in advance the NFT market's dynamics and potential challenges during bear markets. Such foresight assists in preventing rug pulls during market downturns, thereby preserving their reputations [107]. Finally, launchers should also implement criteria for membership to filter out purely speculative participants. For instance, inclusion could be based on additional factors like member recommendations or skill demonstrations, ensuring a diverse and dedicated community.

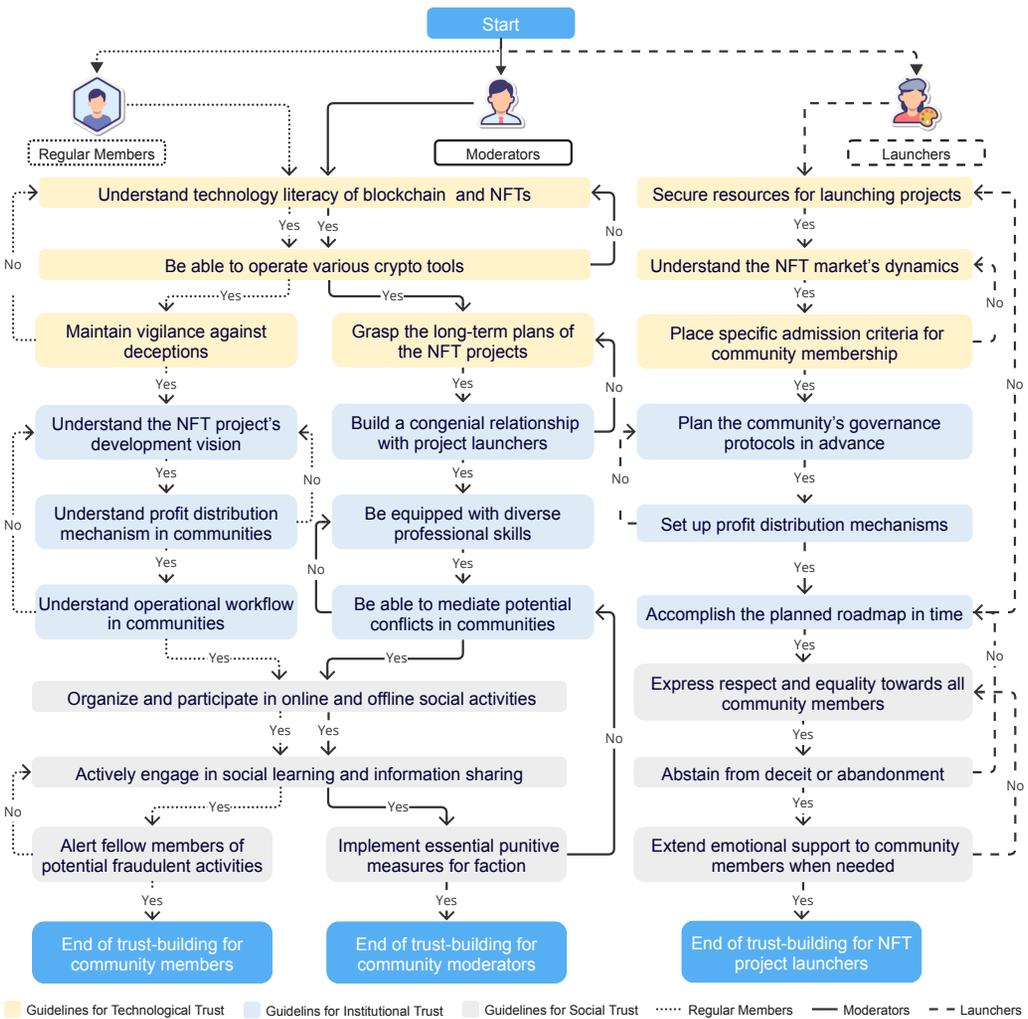

Fig. 4. Action framework for trust building in Chinese NFT communities based on proposed guidelines. The flowchart outlines strategies for regular members, moderators, and launchers to cultivate technological, institutional, and social trust. Participants can selectively employ these strategies during actual interactions, depending on their market roles and current circumstances.

Currently, international NFT communities on Discord use bots and welcome channels to help newcomers grasp NFT basics, while Chinese NFT communities rely on WeChat official accounts. However, these methods can be labor-intensive. To enhance onboarding efficiency, communities could adopt NFT tools with user-friendly interfaces, step-by-step tutorials, and comprehensive educational resource libraries. In addition, platforms should perform thorough due diligence and verify the identities of prospective launchers. This approach reduces the risk of allowing individuals with compliant smart contracts but insufficient resources and accountability to start communities.

*5.2.2 Guidelines for Building Institutional Trust.* To establish institutional trust, we recommend that launchers, moderators, and members adopt complementary yet slightly distinct strategies.

In particular, launchers must meticulously plan the community's governance protocols. Governance protocols could involve determining which collaboration tools to employ, whether to set up a treasury, how democratic voting should be conducted, and the mechanisms for profit distribution when the community reaps benefits. Considerations could extend to using copyright fees, allocating profits from derivatives, and deploying additional economic incentives, such as WeChat red envelopes.

Platforms can further assist launchers in optimizing community management and moderation by integrating Web 3.0 community management tools, such as Charmverse [8]. These tools require members to register using their wallet addresses. When members complete assigned tasks, their wallets automatically receive predetermined bounties, streamlining the reward process and fostering active participation.

Moderators must enforce the governance protocols established by the launchers to strengthen institutional trust. The role of moderators is critical in Chinese NFT communities, where many members face language barriers. They are responsible for the daily management of the community and directly interact with members. Specifically, their duties require establishing a respectful relationship and efficient communication channels with launchers. They should be equipped with diverse professional skills, such as organizing X spaces and guiding newcomers using crypto tools. Moreover, they need to possess a strong sense of justice and empathy to mediate potential conflicts within the community.

Regular members must grasp a comprehensive understanding of the NFT project's development vision, operational workflow, and profit-sharing mechanisms. However, Chinese NFT community members often rely on fragmented information from sources such as X or translations provided by volunteers. To improve access to reliable updates, regular members and local moderators can use collaborative platforms such as Notion to archive crucial information. This strategy helps them maintain a dialectical mindset and establish realistic expectations of potential economic gains.

*5.2.3 Guidelines for Building Social Trust.* Building social trust primarily requires a collaborative effort between moderators and members to foster a harmonious community atmosphere. Typically, moderators and members organize and participate in both online and offline activities to cultivate a sense of belonging and self-efficacy, thereby enhancing social trust among each other. For instance, they can actively engage in social learning and foster a culture of reciprocity by sharing information with their fellow members [114, 138].

Another essential practice for maintaining social trust is vigilance towards suspicious individuals or potential fraudulent activities. Regular members should be encouraged to alert one another upon noticing such incidents. Nevertheless, solely relying on manual screening is inefficient and prone to errors. To enhance efficiency, transaction platforms can integrate automated NFT spam and fraud detection algorithms to assist regular members in identifying suspicious users. Once suspicious entities are identified, moderators can consider revoking the memberships of reported individuals if their wallet addresses match their WeChat IDs. Moreover, they should remain alert to any attempts to create divisions, spread unfounded rumors, or isolate members within the community and be prepared to implement appropriate punitive measures when necessary.

The role of launchers in building social trust stems from their commitment to socio-cultural norms, particularly the Confucian virtues. They are tasked with establishing clear rules to protect community members' *mianzi* by demonstrating basic respect in their interactions [66]. When interpersonal conflicts arise, launchers should be ready to offer emotional support, at the very least, to community members, particularly those experiencing FUD, isolation, or defamation. In these situations, the support and encouragement of the launchers is vital.

## 5.3 Navigating the Potential Pitfalls of Guanxi: Design Implications for Trust Challenges

This subsection presents design implications for platform advancement to address trust challenges in Chinese NFT communities. Issues such as fraud, speculative bubbles, rug pulls, and a pervasive gambling mentality are particularly pronounced in the Chinese context but are also common globally [93, 107, 114, 123]. Therefore, the insights gained from the Chinese experience can be applied by researchers, developers, and designers to enhance trust and user experience in NFT communities worldwide.

*5.3.1 Design Implications for Technological Trust Challenges.* An overreliance on personal relationships hugely challenges technological trust within Chinese NFT communities. The crux of the issue lies in the current inadequacy of crypto tools to provide risk warnings and protect user rights effectively. Compounding this problem is the immutable nature of blockchain transactions, which represents a technical dilemma that the entire NFT ecosystem is grappling with [42, 126].

To address these issues, transaction platforms and social media can enforce existing policies that target platform manipulation and financial scams. This may involve platforms establishing security committees [4] to monitor interaction behaviors, act on complaint records, and flag suspicious links before stakeholders authorize transactions. Further security measures could involve using blockchain-specific decentralized identity solutions to authenticate all parties in a transaction [75, 90]. For instance, the Ethereum Improvement Proposal (EIP)—ERC 6651, introduced in 2023, assigns each NFT its smart contract account or wallet, establishing on-chain identities linked to social network models and effectively mitigating the risk of impersonation scams.[10]

Moreover, integrating escrow services can safeguard against scams and market hype [23, 54]. By holding the buyer's funds until transaction conditions are met, these services introduce a protective buffer for community members, potentially swayed by FOMO or FUD, enhancing overall transaction security and technological trust.

*5.3.2 Design Implications for Institutional Trust Challenges.* Institutional trust is mainly undermined by overly optimistic launchers. With the legal and economic frameworks for NFT project launches still in flux, many community members turn to social media for information, such as project updates in WeChat groups and live broadcasts on X Spaces [31, 114]. Though somewhat effective, this approach can lack objectivity and lead to misjudgments about the intention and competence of launchers. This issue is not unique to China but is a widespread concern across global NFT marketplaces.

Therefore, it is crucial to equip platforms with functionalities that prevent and resolve disputes arising from launchers' mismanagement. For prevention, transaction platforms can incorporate algorithms for risk assessment and anomaly detection in the smart contracts of upcoming NFT projects. Developers might introduce plugins for crypto wallets to audit smart contracts and perform formal verification [11, 94]. These tools can identify security vulnerabilities that attackers might exploit to compromise contracts, steal funds, or disrupt services. For dispute resolution, platforms can leverage arbitration services like Kleros [108] to settle conflicts and deter fraud or abandonment by project launchers. These strategies not only help identify ill-prepared launchers but also provide a fair and reasonable exit mechanism, potentially reducing instances of rug pulls.

*5.3.3 Design Implications for Social Trust Challenges.* Social trust is threatened by members with unrealistic expectations, prioritizing personal interests over community welfare. Compared to the other two dimensions, social trust faces challenges that are more specific to the Chinese context.

---

[10]https://eips.ethereum.org/EIPS/eip-6551

However, the underlying issue of insufficient oversight of inappropriate behavior among community members is prevalent in NFT communities from various societies and cultures [25, 114, 136].

Therefore, we propose implementing reputation systems for active wallet addresses or NFT influencers, which could be integrated into the community website or related social media platforms to mitigate the risks associated with anonymity. Such systems are designed to deter fraudulent activities, foster safer interactions, and thus preserve social trust among community members.

Building on this concept, we could develop anonymous assessment tools as a feature within these reputation systems, allowing users to evaluate the credibility and market influence of their peers. This functionality would assist community members in making informed judgments about the reliability of others. Some participants have reported using similar third-party tools designed to rank user influence, such as NFT Profile Social Rankings, to identify trustworthy community members for potential collaborations.[11] These tools provide valuable data, such as follower count and post frequency from X, as well as project loyalty and overall influence ranking.

Furthermore, we can use smart contracts to automatically monitor community members' transaction behaviors, enabling the introduction of measures to reduce excessive transactions or deliver on promised rewards. This approach can effectively deter short-term speculators, fostering the community's sustainable and healthy growth.

## 6 Limitations and Future Work

Despite comprehensively investigating stakeholders' trust perceptions and relevant obstacles within Chinese NFT communities, our study also presents certain limitations.

First, our study shares common limitations with other qualitative research, including the limited sample size [61] potential researcher biases [52], and the unconfirmed generalizability of our findings [60]. Thus, further research is needed to extend and validate our guidelines and design implications in diverse socio-cultural contexts, which we plan to undertake in future work. Despite these limitations, our findings offer valuable insights into trust dynamics in global NFT and online communities. For example, technological trust challenges such as security concerns and unpredictable pricing, observed in Chinese NFT communities, are also prevalent worldwide [93, 114, 137]. Similarly, social trust issues like dark participation and volatile memberships are common across various social media platforms [82], online gaming communities [98], and online political communities [119]. Therefore, the insights from our study offer a preliminary framework that could be adapted to different social and cultural settings.

Second, although our interview sample covered three of the most active market roles in Chinese NFT communities—namely launchers, moderators, and regular members—these categories do not comprise all stakeholders identified by others [21, 130]. Our sample did not include roles such as social media influencers, investors, and developers. Their exclusion was due to their smaller representation in the total participant count and limited direct interaction with NFT communities. Nonetheless, their attitudes towards trust and the challenges they face could offer valuable insights to validate and supplement the findings of this study. We propose that future researchers cross-validate the perspectives of a more comprehensive stakeholder sampling pool on trust perceptions and relevant challenges.

Third, our study relies solely on textual and narrative data. Information from images and external URLs was disregarded. Thus certain subtle and complex emotions, trust perceptions, and challenges might have been overlooked or neglected.

---

[11]https://www.inspect.xyz/rank/profiles

## 7 Conclusion

This study extends the understanding of trust-building within NFT communities, viewed from sociocultural perspectives. Our two-fold empirical, qualitative study incorporates a content analysis of eight WeChat groups along with 21 semi-structured interviews with stakeholders in Chinese NFT communities. We unveil the impact of traditional Confucian philosophies, deeply embedded in the cultural constructs of *guanxi* and *mianzi*, on shaping trust perceptions amongst community members. Additionally, we identify three dimensions of trust: technological, institutional, and social, which resonate with the classic framework for analyzing trust perceptions in the broader blockchain ecosystem. From this, we distill and classify the challenges associated with building and preserving each dimension of trust within the Chinese cultural context. Guided by our findings, we propose comprehensive, interpersonal, behavior-oriented guidelines for stakeholder trust-building in NFT communities. We also suggest design implications for developers and designers to enhance trust in Chinese NFT communities. Although our research context is culturally specific, the trust-related challenges we have identified reflect broader patterns that could be relevant across global NFT communities and potentially other types of online communities as well. Therefore, the insights from our research can also be applied beyond China, serving as a reference for building trust in settings such as international NFT communities and various online groups.

## Acknowledgments

We extend our sincere gratitude to the Chinese communities of Soulda, Mfers, Goblintown, BoxCat-Planet, Degen Toonz, Mimic Shhans, Nyolings, and Ape Reunion for their support of this research. We are especially grateful to @UsDesci, @Renee7eth (co-founder of PeoplEarth), and @nft66_666 (co-founder of SofaDAO) for their assistance with recruitment and their valuable insights. We also acknowledge the valuable collaboration of the Crypto-Fintech Lab at HKUST. Finally, we express our appreciation to the anonymous reviewers for their constructive feedback.

# A Keyword lists for Content Analysis
## A.1 Keyword list I

Table 5. Keyword list I: Concepts of trust perceptions and influential factors from the 21 extant papers surveyed

| 10 Concepts of Trust Perceptions | | |
|---|---|---|
| **Keywords** | **Synonyms** | **References** |
| Benevolence | Care/Caring/Affinity/Mutual support | Jarvenpaa et al. (1998) [67], Riegelsberger et al. (2005) [100], Gaggioli et al. (2019) [51], Iyer et al. (2020) [66], Shareef et al. (2020) [111], Sharma et al. (2022) [114], Zhang et al. (2023) [138] |
| Credibility | Authenticity/Reputation | Jarvenpaa et al. (1998) [67], Riegelsberger et al. (2005) [100], Porter et al. (2008) [97], Sharma et al. (2022) [114], Zhang et al. (2023) [138] |
| Judgment | Accuracy/Correctness | Jarvenpaa et al. (1998) [67], Sharma et al. (2022) [114], Zhang et al. (2023) [138] |
| Willingness to Cooperate | Openness/Acceptance/Alignment/ Willingness to Share Information | Porter et al. (2008) [97], Gaggioli et al. (2019) [51], Sharma et al. (2022) [114] |
| Loyalty Intentions | - | Porter et al. (2008) [97], Gaggioli et al. (2019) [51] |
| Competence | Knowledge/Expertise/Strength/Ability | Riegelsberger et al. (2005) [100], Mpinganjira. (2018) [89], Gaggioli et al. (2019) [51], Sharma et al. (2022) [114] |
| Predictability | Conformity | Riegelsberger et al. (2005) [100], Shareef et al. (2020) [111] |
| Integrity | Honestly | Jarvenpaa et al. (1998) [67], Gaggioli et al. (2019) [51], Iyer et al. (2020) [66], Shareef et al. (2020) [111], Sharma et al. (2022) [114] |
| Reliability | Dependability | Iyer et al. (2020) [66], Shareef et al. (2020) [111] |
| Dedication | Voluntary Cooperation/Collaboration | Shareef et al. (2020) [111] |
| **17 Concepts of Trust Influential Factors** | | |
| **Keywords** | **Synonyms** | **References** |
| Social Context | Environment Layer (the social, legal and economic contexts) | Gaggioli et al. (2019) [51], Zhang et al. (2023) [138] |
| Perceive a Sense of Respect | - | Porter et al. (2008) [97] |
| Rewards | Incentives | Riegelsberger et al. (2005) [100] |
| Information Usefulness | Communication/Information Sharing | Jarvenpaa et al. (1998) [67], Mpinganjira (2018) [89], Zhang et al. (2023) [138] |
| Community Responsiveness | - | Mpinganjira (2018) [89], Ridings et al., (2002) [99] |
| Consensus | Shared Values/Shared Vision | Porter et al. (2008) [97], Mpinganjira (2018) [89], Shareef et al. (2020) [111], Sharma et al. (2022) [114] |
| Experience | - | Abdul-Rahman et al. (2000) [12], Gaggioli et al. (2019) [51], Sharma et al. (2022) [114] |
| Personal Interactions | Friendship Formation/Connectedness | Iyer et al. (2020) [66] |
| Satisfaction | Fulfilled Expectation/Emotional Reasons | Shareef et al. (2020) [111], Sharma et al. (2022) [114] |
| Active Engagement | Frequency and Pattern of Interaction | Jarvenpaa et al. (1998) [67], Iyer et al. (2020) [66] |
| Transparent Transactions | - | Sas et al. (2017) [106], Zarifis et al. (2022) [136] |
| Perception of Technology's Credibility | Expertise/Knowledge/Perception of Risk/ Perception of Technology's Ease of Use/ Understanding of Blockchain Concept | Riegelsberger et al. (2005) [100], Sharma et al. (2022) [114] |
| Social Embeddedness | Social Practice/Promote Event/Social Interaction/Content & writing Style/ Tone (Communication Style) | Jarvenpaa et al. (1998) [67], Riegelsberger et al. (2005) [100], Sharma et al. (2022) [114], Zhang et al. (2023) [138] |
| Familiarity | Similarity (with the person or group)/ Compatibility | Shareef et al. (2020) [111], Sharma et al. (2022) [114] |
| Security | - | Gaggioli et al. (2019) [51], Das et al. (2022) [42] |
| Subjective norms | Interpersonal Context (attitudes toward risky prospects, betrayal sensitivity, and trustworthiness expectations) | Gaggioli et al. (2019) [51], Shareef et al. (2020) [111] |
| Demographic | - | Zhang et al. (2023) [138] |

## A.2　Keyword list II

Table 6. Keyword list II: Frequent Chinese *jargon* of NFT communities identified from chat threads from eight WeChat groups with an English translation and corresponding explanations.

| Jargons | English Translation | Explanation |
|---|---|---|
| 福報 | Blessing | Often metaphorically refers to the good fortune or luck in obtaining valuable NFTs. |
| 鐮刀 | Sickle | A metaphorical expression in the NFT trading market, referring to deceptive or speculative launchers who use unethical means to take money from users. |
| 掃地板 | Sweeping the Floor | Metaphorically refers to the act of purchasing and holding unsold low-priced NFTs to drive up the floor price of the entire NFT collection project. |
| 二創 | Secondary Creation | Derivative works based on the original NFTs. |
| 牛歸速回 | "Bulls Return Quickly" | A wishful expression hoping for a quick market recovery or the return of a bull market. |
| 加倉 | Add Position | Refers to purchasing more NFTs to increase one's holdings. |
| 接offer | Accept Offer | Accepting a bid or offer for an NFT. |
| FOMO | Fear of Missing Out | Anxiety that an exciting event may currently be happening elsewhere. In the NFT context, it often refers to the fear of missing out on a valuable NFT purchase. |
| 空投 | Airdrop | The action of launchers distributing derivative NFTs, cryptocurrencies, or NFTs of new projects for free to the wallet addresses of community members. |
| 白單 | Whitelist | A whitelist grants registered wallet addresses of community members additional benefits, such as priority in purchasing derivative NFTs or free chances in lotteries, among others. |
| 韭菜 | Leeks | A metaphor for newcomers or unexperienced stakeholders who are often "harvested" by deceptive or speculative launchers in market manipulations. |
| 土狗 | Mutt Projects | A metaphorical expression of inferior NFT projects that have a lower chance of appreciation and are unlikely to become blue-chip assets in the market. |
| FUD | Fear, Uncertainty, and Doubt | Disinformation strategy used in NFT market manipulation to influence perception by disseminating negative, dubious or false information. |
| 仿盤 | Copycat | NFTs that are replicas or imitations of popular or valuable NFTs. |
| 元宇宙 | Metaverse | VR galleries or lands that originated from NFT projects. Occasionally, play-to-earn games with NFTs as pass cards. |
| 家人 | Family | Community members holding NFTs from the same NFT collection projects. |
| 紅包 | Red Envelope | A traditional Chinese gift, now often digital, containing money. In Chinese NFT communities, distributing digital red envelopes in WeChat groups fosters a sense of harmony. |
| 騙子 | Scammer | In NFT communities, scammers trick stakeholders into interactions to steal from their cryptocurrency wallets through phishing websites and identity fraud. |
| 跑路 | Rug and Pull | An operator of a fraudulent business disappearing with the investors' money. |
| 國庫 | Treasury | In DAOs (Decentralized Autonomous Organizations) and some NFT communities, the collective funds or assets that the organization owns. |
| 多簽 | Multisig | A technology used in cryptocurrency wallets that requires multiple signatures for a transaction to occur, improving security. |
| 投票 | Voting | A common mechanism in DAOs where token holders vote on decisions or proposals. |
| 發推 | Tweet | Posting a message on X (formerly Twitter). |

## B Survey Guide in Semi-structured Interview

### 1. Experience Regarding NFT Communities

(1) How many years have you participated in NFT communities?
(2) How many Chinese NFT communities have you participated in?
(3) Which Chinese NFT project communities have you participated in?
(4) Do you also participate in international NFT communities?
   [ ] Yes
   [ ] No

   *If yes, then:*
  (a) How many international NFT communities have you participated in?
  (b) Please name the international NFT communities in which you have participated.
   *If no, then:*
  (a) Please briefly indicate the reason.
(5) How would you describe your level of experience in Chinese NFT communities?
  (a) Expert/Original Gangster (OG)
  (b) Highly Experienced
  (c) Experienced
  (d) Somewhat Experienced
  (e) Novice
(6) How would you define your role in Chinese NFT communities?
  (a) Project Launchers
  (b) Moderators
  (c) Builders / Informal Leaders / Active Members
  (d) Regular Members
  (e) Others (please specify)

### 2. Perceptions and Expectations for NFT Communities

(1) What are your motivations for participating in Chinese NFT communities?
  (a) Have fun with other holders.
  (b) Quickly obtain the latest project updates and market information.
  (c) Find collaborators or potential users for other business activities.
  (d) Satisfy a sense of belonging and identity.
  (e) Better contribute to invested NFT projects.
  (f) Get incentives.
  (g) Others (please specify).
(2) Do you believe in the long-term development of the NFT ecosystem?
  (a) Very pessimistic
  (b) Pessimistic
  (c) Neutral
  (d) Optimistic
  (e) Very optimistic
(3) How much do you trust the Chinese NFT communities you participate in?
  (a) Very distrustful
  (b) Somewhat distrustful
  (c) Neutral
  (d) Somewhat trustful
  (e) Very trustful
(4) How much do you trust the moderators and launchers of the Chinese NFT communities?

   (a) Very distrustful
   (b) Somewhat distrustful
   (c) Neutral
   (d) Somewhat trustful
   (e) Very trustful
(5) How much do you trust fellow members within the Chinese NFT communities?
   (a) Very distrustful
   (b) Somewhat distrustful
   (c) Neutral
   (d) Somewhat trustful
   (e) Very trustful

### 3. Demographics Information
(1) What is your age range?
   (a) 18-24
   (b) 25-29
   (c) 30-34
   (d) 35 and above
(2) What is your gender?
   (a) Male
   (b) Female
   (c) Non-binary/Third gender
   (d) Prefer not to say
(3) Which city are you currently located in?
(4) What is your current occupation?
(5) What is your major and educational background?

# C  Details for Five Confucian Virtues

Table 7.  Illustration of the *Five Confucian Virtues* and the separate frequency of their occurrence within the processed data from content analysis and semi-structured interviews.

| Five Confucian Virtues | | Frequencies | |
|---|---|---|---|
| | | CA | SSI |
| 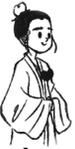 **Benevolence (仁)** | **Benevolence** is one of the core values of Confucian thought, emphasizing care and compassion for others. In Confucian thought, a benevolent person cares about the well-being of others, empathize with the feelings of others, and considers the interests of others. | 24.7% | 20.09% |
| 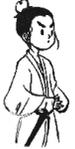 **Righteousness (義)** | **Righteousness** is another key concept in Confucian ethics, requiring people to make morally correct choices and fulfill their responsibilities. Righteousness emphasizes an individual's roles and responsibilities within the family, society, and the state. | 23.48% | 13.67% |
| 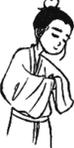 **Propriety (禮)** | **Propriety** is a concept in Confucian thought regarding behavioral norms and social etiquette. Propriety aims to maintain social order and harmony, requiring people to follow certain rules and etiquette to express respect. | 15.24% | 21.8% |
| 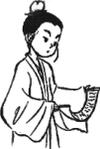 **Wisdom (智)** | **Wisdom** is an aspect of Confucian thought that emphasizes the wisdom and judgment that people should possess. In Confucian ethics, wisdom is manifested as the ability to discern right from wrong, make wise decisions, and understand knowledge and experience. | 19.82% | 15.81% |
| 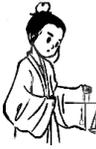 **Integrity (誠)** | **Integrity** is an aspect of Confucian ethics that emphasizes honesty and loyalty. Integrity requires people to be sincere, and reliable in their conduct to gain the credibility of others. | 16.76% | 28.63% |